\documentclass[chicago]{emulateapj}
\usepackage{graphicx}

\slugcomment{Accepted by The Astrophysical Journal}

\shorttitle{Warm H$_2$ Emission And Ram-Pressure Stripping}

\begin{document}

\newcommand{\etal}{{\it et al.}}
\newcommand{\htwo}{H$_2$ }
\newcommand{\ha}{H$\alpha$ }
\newcommand{\mic}{$\mu$m }
\newcommand{\ms}{M$_\odot$ }
\newcommand{\msyr}{M$_\odot$ yr$^{-1}$ }

\title{Tracing Ram-Pressure Stripping with Warm Molecular Hydrogen Emission}

\author{Suresh Sivanandam\altaffilmark{1,2}, Marcia J. Rieke\altaffilmark{2}, and George H. Rieke\altaffilmark{2}}

\altaffiltext{1}{Dunlap Fellow, Dunlap Institute for Astronomy and Astrophysics, University of Toronto, Rm 101, 50 St. George St, Toronto, ON, Canada M5S 3H4; sivanandam@di.utoronto.ca}
\altaffiltext{2}{Steward Observatory, University of Arizona, 933 North Cherry Ave, Tucson, AZ, USA 85721}

\begin{abstract}
We use the \emph{Spitzer} Infrared Spectrograph (IRS) to study four infalling cluster galaxies with signatures of on-going ram-pressure stripping. H$_2$ emission is detected in all four; two show extraplanar H$_2$ emission.  The emission usually has a warm (T $\sim$ $115 - 160$K) and a hot (T $\sim$ 400 $-$ 600K) component that is approximately two orders of magnitude less massive than the warm one. The warm component column densities are typically $10^{19} - 10^{20}$ cm$^{-2}$ with masses of $10^6 - 10^8 M_\odot$. The warm H$_2$ is anomalously bright compared with normal star-forming galaxies and therefore may be excited by ram-pressure. In the case of CGCG 97-073, the H$_2$ is offset from the majority of star formation along the direction of the galaxy's motion in the cluster, suggesting it is forming in the ram-pressure wake of the galaxy. Another galaxy, NGC 4522, exhibits a warm H$_2$ tail approximately 4 kpc in length. These results support the hypothesis that H$_2$ within these galaxies is shock-heated from the interaction with the intracluster medium. Stripping of dust is also a common feature of the galaxies. For NGC 4522, where the distribution of dust at 8 $\mu$m is well resolved, knots and ripples demonstrate the turbulent nature of the stripping process. The H$\alpha$ and 24 $\mu$m luminosities show that most of the galaxies have star formation rates comparable to similar mass counterparts in the field. Finally, we suggest a possible evolutionary sequence primarily related to the strength of ram-pressure a galaxy experiences to explain the varied results observed in our sample.

\end{abstract}


\section{Introduction}
Ram-pressure stripping is thought to transform infalling galaxies in the cluster environment \citep{gunn72} and may explain the distinctly different properties of cluster galaxies when compared to the field.  The traditional paradigm of stripping confined to just atomic gas is being challenged. Observations show that molecular gas can be stripped from a cluster galaxy and extend to well beyond the tidal radius of the galaxy \citep{sivanandam10}. Other observations show star forming trails emanating from infalling galaxies in clusters \citep{cortese07,sun10,sivanandam10}. While originally thought to be fairly rare \citep{sun10}, recent deep narrowband \ha and \emph{GALEX} UV observations in the Coma cluster show at least a dozen galaxies with long ($\sim 20-100$ kpc) \ha trails \citep{yagi10} and UV trails \citep{smith10} all generally pointing away from the cluster centre, indicating that ram-pressure can strip gas, and stars can form well beyond the gravitational well of the stripped galaxy. 
\par
The traditional view is that ram-pressure does not have sufficient strength to remove significant amounts of molecular gas from galaxies, based on the observation that cluster galaxies typically show little correlation between deficiencies in atomic (HI) and molecular gas \citep{boselli06}. The removal of just the HI gas, which in turn no longer replenishes the \htwo reservoir, will impact the star forming properties of galaxies only slowly, since the \htwo gas consumption time scale in nearby disk galaxies is approximately 2 Gyr \citep{bigiel11}. The observations of stripped molecular gas suggest that the suppression process may operate more quickly. Other studies also suggest that the molecular component is affected by ram pressure, because there is a correlation between \htwo deficiency and galaxies that have lost HI within their optical disks \citep{fumagalli09}. Furthermore, Herschel observations of cluster galaxies reveal truncated disks, indicating erosion of their molecular components \citep{cortese10,corbelli12}. The rapid removal of molecular gas would have a relatively quick impact on the star-forming properties of galaxies. This possibility would agree with other studies \citep[e.g.,][]{poggianti99, bai07} that infer relatively rapid quenching of star formation in clusters.
\par
A more complete view is required to understand fully the role ram-pressure stripping plays in the evolution of a cluster galaxy. The discovery of a galaxy group scale shock in Stephan's Quintet \citep{appleton06, cluver10} glowing in ground-state rotational lines of warm molecular hydrogen (H$_2$) pointed toward a new method to detect ram-pressure stripped molecular hydrogen, and to study the shock associated with the interaction between the intracluster medium (ICM) and interstellar medium (ISM) of a galaxy experiencing ram-pressure stripping. \cite{cluver10} used \emph{Spitzer}'s infrared spectrograph to generate a large-area spectral map of this shock; they find a large amount of \htwo ($\sim5\times10^8$ M$_\odot$) within the main shock region where ground-state rotational lines of warm \htwo are the main coolant. Unexpectedly, considerable CO emission, a tracer of cold H$_2$, was discovered within this shock \citep{guillard12}, indicating that there is a significant cold gas reservoir in the intragroup medium. The group-wide shock is thought to be produced by the collision of an intruder galaxy with the intragroup medium at a velocity of $\sim1000$ km s$^{-1}$ \citep{cluver10}. The collision velocity is similar to the relative velocity between the ISM of an infalling galaxy and the ICM of a dense cluster, which motivated our survey for similar features of shocked \htwo emission in candidate galaxies in nearby clusters. The \htwo emission may serve as a direct tracer of the removal of molecular gas in these galaxies, which may immediately impact the star forming properties of these systems. 
\par
We have searched for warm molecular hydrogen (H$_2$) emission from four galaxies that show strong signs of on-going ram-pressure stripping. ESO 137-001, an infalling spiral in the rich cluster Abell 3627, was the first galaxy observed by our program and was also the best candidate for detecting this emission, as it had spectacular X-ray and H$\alpha$ tails extending 70 and 40 kpc, respectively \citep{sun07}. Through {\it Spitzer} Infrared Spectrograph (IRS) imaging spectroscopy, we detected a striking warm H$_2$ tail (T$\sim 130-160$K) extending at least 20 kpc  from this galaxy \citep[][ hereafter referred to as Paper 1]{sivanandam10}. The H$_2$ tail was found to be co-aligned with the X-ray and H$\alpha$ tails, and extended farther than the tidal radius of the galaxy.  The tail included most of the extraplanar star forming regions, which led to our conclusion that the H$_2$ being stripped from the galaxy fostered star formation. The most tantalizing discovery was that the warm H$_2$ emission was not powered by star formation, which is the most common excitation mechanism. Similar signatures of unusually strong warm H$_2$ emission have been detected in cool-core clusters in their central regions within their brightest cluster galaxies \citep{egami06,donahue11} and along extended filaments \citep{johnstone07}. Within this cooling flow environment, multiple heating mechanisms for H$_2$ have been suggested including cosmic ray heating, conduction, and dissipative magnetohydrodynamic (MHD) waves \cite{ferland08,ferland09}. Other potential mechanisms have also been discussed such as X-ray dissociation regions (XDRs) \citep{maloney96}. Our previous work suggested that the H$_2$ is being dissociated by a shock produced by the interaction between the ICM and galactic interstellar medium and is reforming in the post-shock region in an excited state. Recently \cite{wong14} also detected similar enhanced \htwo emission, albeit at a lower level, in four ram-pressure stripped Virgo galaxies and they also favour shock excitation as the explanation for the anomalous \htwo emission. However, these alternative heating mechanisms have not been definitely ruled out.
\par
We explore these possibilities further in our expanded sample of three more galaxies that have clear signs of ongoing ram-pressure stripping. CGCG 97-073 (hereafter referred to as 97073) is an irregular galaxy with a 50 kpc long H$\alpha$ tail \citep{gavazzi01} in another rich cluster, Abell 1367. NGC 4522 is a type SBcd spiral in the Virgo cluster with arc shaped H$\alpha$ emission and with prominent extraplanar star formation, a truncated H$\alpha$ disk \citep{kenney99}, and a truncated and displaced HI distribution \citep{kenney04}. Finally, NGC1427A is a dwarf irregular galaxy in the Fornax cluster with prominent arc-shaped morphology and star formation along one edge \citep{georgiev06}, which is likely explained by the interaction with the Fornax's ICM \citep{chaname00}. Our detecting and characterizing warm \htwo and extended emission by dust from these galaxies provides a more complete picture of how common molecular gas stripping is in different cluster environments, and how ram-pressure impacts infalling galaxies. 
\par
The paper is organized as follows. In Section 2, we present the sources of our data and the methodology used to analyze the imaging and spectroscopic data. In Section 3, we present the results from our data reduction, and characterize the properties of the warm \htwo and dust and determine the star-forming properties of the galaxies. In Section 4, we place the star formation rates of the sample galaxies in a broader context and discuss the possible excitation mechanisms for H$_2$, including how effectively excess warm \htwo may trace ram-pressure stripping. We also provide a possible explanation for the variety of results among these galaxies. Finally in Section 5, we list our conclusions.  For computing distances, we adopt the concordance cosmological model ($\Omega_\Lambda = 0.73,$ $\Omega_m = 0.27,$ and $H_0 = 71$ km s$^{-1}$ Mpc$^{-1}$). All reported errors in this work are quoted at the 1$\sigma$ level.

\section{Observations and Data Reduction}
To detect and characterize the  warm molecular hydrogen in our sample of galaxies, we carried out a {\it Spitzer} GTO program 50213 (PI: G. Rieke), in which we spectrally mapped the galaxies with {\it Spitzer}'s infrared spectrograph (IRS) and imaged the galaxies with the infrared array camera (IRAC). We also compiled an ancillary dataset consisting of archival {\it Hubble}, GOLDMine \citep{gavazzi03}, and {\it Spitzer} data. This heterogenous dataset consists of optical-band {\it HST} images, H$\alpha$ images, additional \emph{Spitzer} IRAC images, and additional IRS spectral mapping data. 

\subsection{IRAC Imaging}
We obtained data for each galaxy in all four (3.6, 4.5, 5.7, and 8 $\mu m$) IRAC channels \citep{fazio04}. IRAC imaging of 97073 and NGC 1427A was carried out as part of our {\it Spitzer} GTO program. IRAC data for NGC 4522 were obtained from the {\it Spitzer} Science Archive, observed as part of program 30945 (PI. J. Kenney).  We constructed IRAC mosaics of the galaxy with MOPEX (version 18.3.1) using the Basic Calibrated Data (BCD) from the Spitzer pipeline (version S18.7) as input and following the {\it Spitzer} Science Center (SSC) IRAC reduction cookbook. To reduce banding artifacts, corrected BCDs were used as input for the construction of 8 $\mu$m mosaics. Bad pixel rejection was achieved through dithering, median filtering, and masking of known bad pixels using pixel masks provided by the SSC.
\par
We also construct 8$\mu$m excess images for each galaxy to study its star-forming properties using the same method outlined in Paper 1. Excess 8 $\mu$m flux in galaxies is often produced by warm dust rich in aromatic hydrocarbons, which is associated with star formation regions \citep{calzetti07}. Our technique subtracts the stellar continuum from the 8 $\mu$m image by subtracting a resolution-matched, flux-scaled, and well-registered 3.6 $\mu$m image. We verify the effectiveness of our method by ensuring the foreground stars in our field are adequately subtracted in the final 8 $\mu$m excess image.

\subsection{HST Imaging}

We obtained \emph{HST} Advanced Camera for Surveys (ACS) archival data for  NGC 1427A (Proposal ID: 9689; PI: M. Gregg) and NGC 4522 (Proposal ID: 9773; PI: J. Kenney). We used the F660N (narrowband H$\alpha$) and F775W (SDSS I-band) data for NGC 1427A and F435W (B-band) and F814 (I-band) data for NGC 4522. We obtained the pipeline-reduced multidrizzled data for our analysis. To construct the \ha image for NGC 1427A, we used the F660N and F775W filter data. The pipeline reduced data for each filter were already properly registered as part of the pipeline processing. We simply scaled the F775W data and subtracted it from the F660N data to remove the stellar contribution from the narrowband image. We verified the success of our subtraction by ensuring the foreground stars in the field were adequately subtracted. Due to the poor signal-to-noise of the subtracted data, we binned and smoothed the data to improve the visibility of \ha emitting regions. Because of the finite width of the narrowband filter, there will be some contamination from [NII] lines that are close to the \ha wavelength.

\subsection{IRS Spectral Mapping}

\begin{figure}
\epsscale{1.15}
\plotone{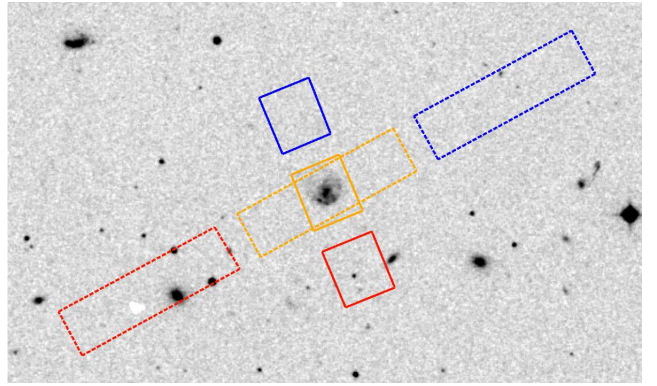}
\caption{6$\arcmin\times$10$\arcmin$ map digitized sky survey (DSS) image of CGCG 97-073 with {\it Spitzer} IRS pointings overlaid. The solid and dashed boxes represent the short-low (SL) and long-low (LL) spectrograph pointings, respectively. The regions with first and second order coverage are shown by the red and blue boxes, respectively. These regions are used for background subtraction. The orange regions have both 1st and 2nd order coverage, and the 2D spectrum of the source is extracted from these regions. \label{97073pnt}}
\end{figure}

\begin{figure}
\epsscale{1.15}
\plotone{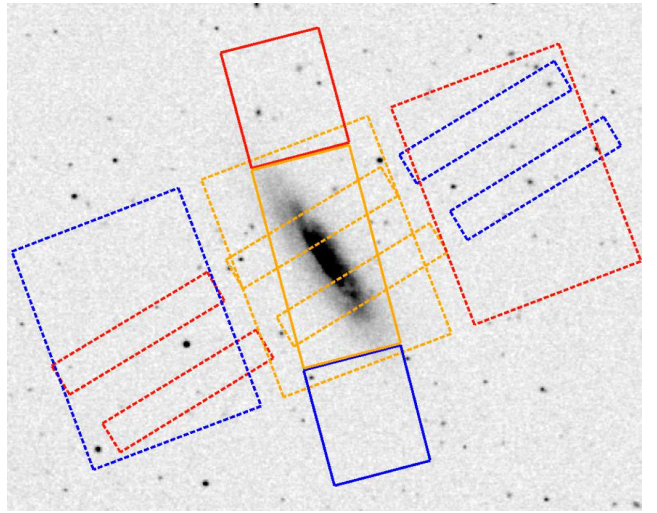}
\caption{8$\arcmin\times$10$\arcmin$ DSS image of NGC 4522. The solid and dashed boxes represent the short-low (SL) and long-low (LL) spectrograph pointings, respectively. The regions with first and second order coverage are shown by the red and blue boxes, respectively. These regions are used for background subtraction. The orange regions have both 1st and 2nd order coverage, and the 2D spectrum of the source is extracted from these regions. There are three sets of LL pointings shown in the image. The larger map provides a complete coverage of the galaxy and is constructed from our own observations.  The two smaller maps, which are centered on regions with known signatures of ram-pressure stripping, are derived from archival data. \label{n4522pnt}
}
\end{figure}

\begin{figure}
\epsscale{1.1}
\plotone{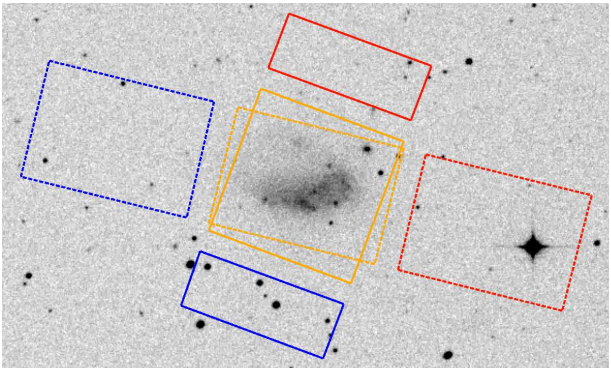}
\caption{6$\arcmin\times$10$\arcmin$ map digitized sky survey (DSS) image of NGC 1427A with {\it Spitzer} IRS pointings overlaid. The solid and dashed boxes represent the short-low (SL) and long-low (LL) spectrograph pointings, respectively. The regions with first and second order coverage are shown by the red and blue boxes, respectively. These regions are used for background subtraction. The first order LL background is partially dominated by a bright star. The slit pointings that overlap the star position are removed from the final background stack. The orange regions have both 1st and 2nd order coverage, and the 2D spectrum of the source is extracted from these regions. \label{n1427apnt}}
\end{figure}

Our IRS observations used both the short-low (SL) and long-low (LL) IRS modules that together span 5.3 to 38.0 $\mu m$ \citep{houck04}. The complete wavelength range allows us to observe several ground vibrational state H$_2$ rotational lines, specifically the $\nu$=0-0 S(0) thru S(7) transitions, that can be used to characterize the thermodynamic properties of warm H$_2$. The spectral mapping techniques and data analysis methods are discussed in Paper 1. We use the most recent IRS pipeline-reduced BCD files (version S18.7) and {\it CUBISM} \citep[version 1.7;][]{smith07a}  to construct our spectral cubes. Bad pixels are eliminated visually using the same method described in Paper 1, and the initial background subtraction is carried out by subtracting the background measured with the outrigger pointings. However, additional background subtraction was required for some observations because the outrigger pointing background subtraction did not remove all of the flux due to background gradients. In these cases, we generated a background spectrum for each spectral map by averaging the spectra in source-free regions. This background was then subtracted, and the error of the background spectrum was included in the final spectral map flux uncertainties. For the case of NGC 4522, we supplement our analyses with archival LL spectrograph data from another {\it Spitzer} program (Proposal ID: 50819; PI: J. Kenney). We use the same methods of background subtraction and bad pixel rejection for analyzing the archival dataset. The coverage maps for the two different sets of observations for NGC 4522 are quite different. Our IRS observations cover the entire galaxy, but the archival observations are deeper and focus on two specific locations where ram-pressure is thought to be stripping significant amounts of gas. The archival observations have approximately two to three times more integration time per sky pixel than our observations ($\sim 90$ s versus $250$ s for LL2 and $150$ s versus $250$ s for LL1) and use a similar mapping strategy. 
The IRS coverage maps for 97073, NGC 4522, and NGC 1427A are shown in Figures \ref{97073pnt}, \ref{n4522pnt}, and \ref{n1427apnt}. The coverage map for ESO 137-001 has already been presented in Paper 1. 

\par
The spectra presented in this paper are first extracted using a custom program that sums the spectra and root-sum-squares the errors within a specified region for each order from the SL and LL spectrographs. The code then stitches the spectra from all orders together without rescaling. The flux values within overlapping wavelength regions are interleaved. The stitched spectrum is then shifted to rest wavelengths using the optical redshift for the galaxy found in the NASA Extragalactic Database (NED). The line fluxes are derived by fitting the extracted spectrum using \emph{PAHFIT} \citep{smith07b}. This routine fits for all \htwo and fine structure lines, aromatic features, and dust continuum within the wavelengths probed by IRS. We typically do not include fitting for extinction except for the case of NGC 4522 where there is a significant amount of dust, as evidenced by the 10\mic silicate absorption feature in the spectra. Strong extinction can significantly change the fluxes of S(1) thru S(7) lines. The dust extinction fitting is discussed in further detail below. The \htwo line fluxes and 1$\sigma$ errors used in our analyses of the \htwo population are the values reported by the \emph{PAHFIT} routine. To determine the contribution of PAH complexes to the spectrum, we used \emph{PAHFIT}'s \emph{pahfit\_main\_feature\_power} routine to accurately determine the errors of correlated features. The H$_2$, PAH feature, and fine structure line fluxes are presented in Tables \ref{h2flux}, \ref{pahflux}, and \ref{fineflux}, respectively. 

\begin{deluxetable*}{lccccccccc}
\tablewidth{0pt}
\small
\centering
\tablecolumns{10}
\tablecaption{Measured H$_2$ Rotational Line Fluxes \label{h2flux}}
\tablehead{
 & & & & & H$_2\:\:0-0$& & & & \\
\colhead{Galaxy} & \colhead{Region} & \colhead{S(0)} & \colhead{S(1)} & \colhead{S(2)} & \colhead{S(3)} & \colhead{S(4)} & \colhead{S(5)} & \colhead{S(6)} & \colhead{S(7)} \\
 & & & & & ($10^{-17}$ W m$^{-2}$)& & & &
}
\startdata
ESO 137-001\tablenotemark{a} & Nuclear & $0.259\pm0.087$ & $1.47\pm0.02$ & $1.02\pm0.16$ & $2.99\pm0.33$ & $<1.68$ & $<2.24$ & $<1.72$ & $2.00\pm0.62$ \\
 & Tail (SL/LL)\tablenotemark{b} & $0.23\pm0.01$ & $0.92\pm0.01$ & $0.42\pm0.08$ & $1.20\pm0.11$ & $<0.59$ & $<1.04$ & $<0.74$ & $1.03\pm0.26$ \\
 & Far Tail (LL-only)\tablenotemark{b} & $0.21\pm0.02$ & $0.80\pm0.09$ & ... & ... & ... & ... & ... & ... \\
CGCG 97-073 & Full & $1.62\pm0.03$ & $5.16\pm0.07$ & $1.79\pm0.56$ & $7.15\pm0.89$ & $<6.36$ & $<1.23\times10^3$ & $<3.96$ & $<9.30$ \\
 & Tail & $0.50\pm0.01$ & $1.43\pm0.03$ & $0.68\pm0.22$ & $2.03\pm0.28$ & $<1.91$ & $<1.69\times10^2$ & $<0.65$ & $<4.92$ \\
 NGC 4522 & Central & $2.92\pm0.09$ & $8.82\pm0.30$ & $3.50\pm1.02$ & $8.23\pm1.86$ & $< 8.28$ & $<15.5$ & $<13.1$ & $<12.3$ \\
 & NE & $0.64\pm0.07$ & $2.00\pm0.14$ & $<2.40$& $<4.47$ & $<6.39$ & $<10.1$ & $<8.58$ & $13.0\pm3.8$ \\ 
 & SW & $0.87\pm0.04$ & $1.83\pm0.07$ & $<2.24$ & $3.68\pm1.12$ & $<5.76$ & $<11.4$ & $<9.69$ & $<10.5$ \\
NGC 1427A & Galactic & $1.16\pm0.08$& $<0.50$ & $<4.41$ & $<4.89$ & $<16.4$ & $<16.8$ & $<14.5$ & $15.0\pm4.3$ \\
& High Flux & $0.81\pm0.05$  & $<0.34$ & $<2.24$ & $<3.09$ & $<8.52$ & $<10.3$ & $<10.1$ & $<7.95$
\enddata
\tablenotetext{a}{See \cite{sivanandam10} for a description of regions.}
\tablenotetext{b}{The H$_2$ fluxes were obtained from \cite{sivanandam10}.}
\tablecomments{All upper limits are 3$\sigma.$}
\end{deluxetable*}

\begin{deluxetable*}{lccccc}
\tablewidth{0pt}
\small
\tablecolumns{5}
\tablecaption{PAH Feature Fluxes \label{pahflux}}
\tablehead{
\colhead{Galaxy} & \colhead{Region} & \colhead{PAH 6.2 $\mu$m} & \colhead{PAH 7.7 $\mu$m} &\colhead{PAH 11.3 $\mu$m} \\
 & & & ($10^{-17}$ W m$^{-2}$)& 
}
\startdata
ESO 137-001\tablenotemark{a} & Nuclear & $65.9\pm1.8$ & $215\pm7$ & $58.1\pm0.9$ \\
	& Tail (SL/LL) & $2.22\pm0.73$ & $21.1\pm2.1$ & $3.43\pm0.29$ \\
CGCG 97-073 & Full & $37.4\pm4.5$ & $80.2\pm14.1$ & $27.6\pm1.6$ \\
 & Tail & $<4.48$ & $<6.91$ & $1.86\pm0.35$  \\
NGC 4522 & Central & $368\pm12$ & $1230\pm40$ & $369\pm7$ \\
 & NE & $<13.5$ & $22.6\pm7.2$ & $41.1\pm3.2$ \\
 & SW &$<15.5$ & $<34.0$ & $45.7\pm3.3$ \\
NGC 1427A & Galactic & $39.1\pm11.1$ & $<48.9$ & $40.3\pm4.1$ \\
& High Flux & $20.6\pm8.0$ & $<23.8$ & $23.6\pm2.2$

\enddata
\tablenotetext{a}{Refer to \cite{sivanandam10} for descriptions of regions.}
\tablecomments{All upper limits are 2$\sigma.$}
\end{deluxetable*}

\begin{deluxetable*}{lccccccc}
\tablewidth{0pt}
\small
\tablecolumns{12}
\tablecaption{Fine Structure Line Fluxes \label{fineflux}}
\tablehead{
\colhead{Galaxy} & \colhead{Region} & \colhead{[Ne II]} & \colhead{[Ne III]} &\colhead{[S III]} & \colhead{[OIV],[FeII]\tablenotemark{a}} &\colhead{[S III]} & \colhead{[Si II]} \\
& & 12.8 $\mu$m & 15.6 $\mu$m & 18.7 $\mu$m & 25.9, 26.0 $\mu$m & 33.5 $\mu$m & 34.8 $\mu$m \\ 
 & & & &($10^{-17}$ W m$^{-2}$)& & &
}
\startdata
ESO 137-001 & Nuclear & $9.80\pm0.24$ & $1.47\pm0.02$ & $4.32\pm0.04$ & $0.45\pm0.01$ & $5.70\pm0.02$ & $8.70\pm0.03$ \\
 & Tail (SL/LL)\tablenotemark{b} &  $1.17 \pm 0.11$ & $0.25 \pm 0.01$ & $0.30 \pm 0.01$ & $0.14 \pm 0.01$ & $0.87 \pm 0.02$ & $1.16 \pm 0.02$ \\
 & Far Tail (LL-only)\tablenotemark{b} & ... & $<0.21$ & $0.17\pm0.04$ & ... & $0.18\pm0.06$ & $0.57\pm0.08$ \\
CGCG 97-073 & Full & $4.05\pm0.77$ & $2.76\pm0.06$ & $3.31\pm0.10$ & $0.32\pm0.03$ &$4.90\pm0.04$ & $5.46\pm0.05$ \\
 & Tail & $<0.56$ & $<0.09$ & $<0.06$ & $0.08\pm0.01$ &$0.45\pm0.02$& $0.60\pm0.02$\\
 NGC 4522 & Central & $30.3\pm1.46$ & $5.56\pm0.19$ & $12.7\pm0.2$ & $4.50\pm0.21$ & $15.0\pm0.2$ & $26.4\pm0.2$\\ 
 & NE & $<3.12$ & $0.66\pm0.17$ & $0.58\pm0.12$ & $0.62\pm0.15$ & $1.80\pm0.13$& $2.80\pm0.16$ \\
 & SW & $<2.26$& $0.91\pm0.06$ & $0.56\pm0.08$ & $0.14\pm0.04$ & $1.66\pm0.06$ & $2.25\pm0.08$ \\
NGC 1427A & Galactic & $<3.63$ & $2.01\pm0.21$ & $2.30\pm0.16$ & $<8.36$ &$0.99\pm0.19$ & $4.77\pm0.26$ \\ 
& High Flux & $<2.42$ & $0.78\pm0.10$ & $1.68\pm0.11$ & $<18.9$ & $<0.38$ & $1.88\pm0.14$
\enddata
\tablenotetext{a}{Blended feature.}
\tablenotetext{b}{Flux values obtained from \cite{sivanandam10}.}
\tablecomments{All upper limits are 3$\sigma.$}
\end{deluxetable*}

\par
Two different types of H$_2$ images are generated for our analyses. The first  does not employ any continuum subtraction. It is generated by summing the rest wavelength spectra within three wavelength bins centered on the 17.035 $\mu$m H$_2$ ($0-0$ $J=3-1$ S(1)) transition. The second type of H$_2$ image is a continuum-subtracted version of the first. The continuum is estimated by generating two adjacent wavelength bins of the same size to the one already generated for the H$_2$ image and then averaging the flux between the two background bins. This method works well for fairly strong H$_2$ features without significant 17 \mic aromatic features, but it is more difficult to detect faint H$_2$ emission in the continuum-subtracted images due to the additional noise introduced by the subtraction.  
\par
We carry out an additional analysis step for NGC 4522 to combine our LL and archival datasets. For generating the 17 \mic images, we coadded the two images generated for each set of observations using the following method: First, we registered and interpolated the image generated from the deeper observations to match the pixel grid of the shallower observations. Second, in regions of coverage overlap, we calculated a weighted average of the two different datasets using the inverse variance of each dataset as weights. Third, in areas without any overlap, we kept the original value associated with the dataset that covered the area. For the generation of spectra, if the chosen extraction region fell entirely within the region where the deeper observations were carried out, we extracted the spectra from just the deeper observations. Otherwise, we use the data from our shallower map.  
\par
We also construct 24 \mic images that mimic what would be observed by the \emph{Spitzer} MIPS imager. We compute a 24 \mic flux with the IRS data for each sky pixel by using the MIPS 24 \mic transmission curve as weights. 

\section{Results}
We present our results for the four galaxies below. In the case of ESO 137-001, we only present new results that were not discussed in Paper 1. This section is organized into two parts: First, we  characterize the warm \htwo emission in our galaxies. We detect significant warm H$_2$ emission in 97073, NGC1427A, and NGC 4522; however in the case of NGC 1427A, we have only sufficient information to place a lower limit on the warm \htwo mass. For the other two galaxies, we present the thermodynamic properties of the observed \htwo and quantify the amount of emitting gas by fitting one or two-temperature models to the excitation diagrams for \htwo within each galaxy and nearby regions of interest. Second, we present the star forming properties of these galaxies. We examine the spatial distribution of warm dust both at 8 \mic and 24 \mic  for signs of ram-pressure induced star formation and determine its correspondence with \ha emission. We measure properties such as star formation rates within the galaxies and compare them to a typical galaxy of the same stellar mass in the field. The four galaxies as an ensemble have star forming properties consistent with their being drawn from the field population. At most, only one of them shows any hint of quenching. It appears that ram-pressure stripping has not significantly impacted star formation in these galaxies.

\subsection{Characterization of Warm \htwo}
\subsubsection{CGCG 97-073}
\begin{figure*}
\epsscale{0.8}
\plotone{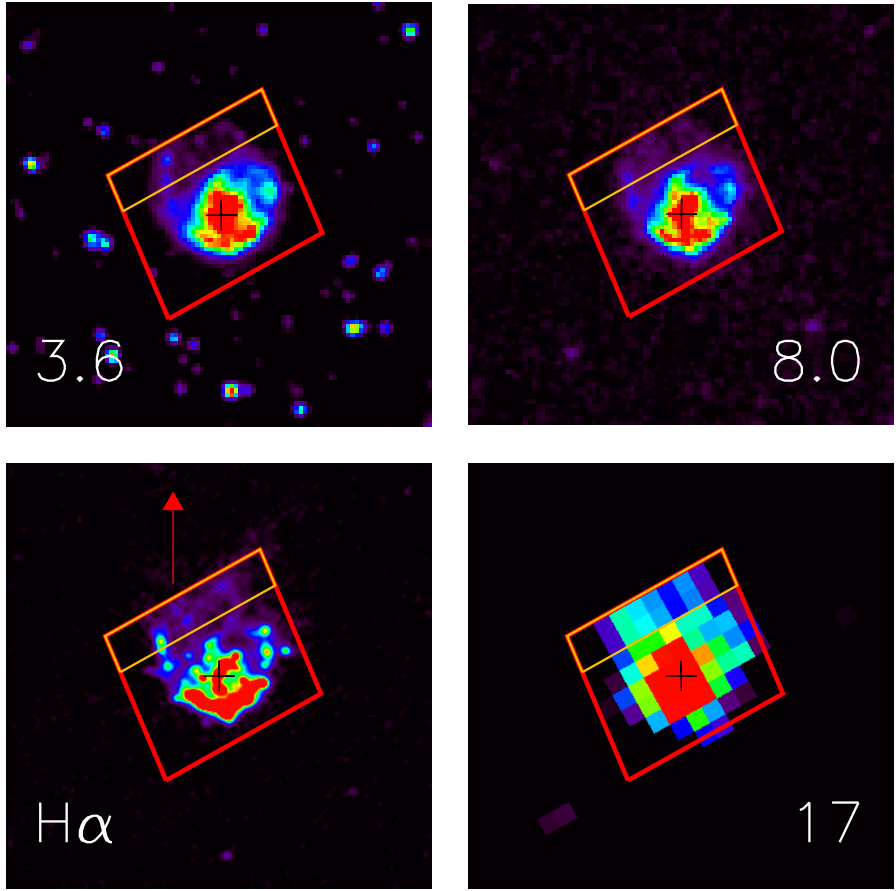}
\caption{ \footnotesize 2$\arcmin\times$2$\arcmin$ images of CGCG 97-073 at 3.6 $\mu$m, 8 $\mu$m, H$\alpha$ (GOLDMine), and rest 17.035 $\mu$m (H$_2$ 0-0 S(1) transition) wavelengths. North is up and East is left. The H$\alpha$ image is smoothed to 1.2$\arcsec$ resolution. The colors follow the visible spectrum where blue represents the faintest emission and red the brightest. The center of the galaxy is show by the black cross. The red and orange boxes represent spectral extraction regions. The red box represents the full galaxy extraction region while the orange one represents the tail extraction region. The red arrow represents the approximate location of the H$\alpha$ tail and its orientation with respect to the galaxy \citep{gavazzi01}. As noted by \cite{gavazzi01}, there is an arc-like feature present in the H$\alpha$ image that is on the side of the galaxy that is directly opposite to the tail. This feature is also observed at 8 $\mu$m, which means it is star formation associated with the ISM/ICM interaction. There is possibly significant H$_2$ emission throughout the galaxy as indicated by the 17 \mic image, including the northern edge of the galaxy, which we consider as the tail, where there is very little stellar emission as shown by the 3.6 $\mu$m image. \label{97073img}}
\end{figure*}

97073 is a galaxy within Abell 1367 that has a measured optical redshift of 0.024267 as reported by the NASA Extragalactic Database (NED). Abell 1367, itself, has a redshift of 0.022, also obtained from NED, which we use to calculate the approximate luminosity distance to the galaxy of 96 Mpc. The galaxy is located at a projected distance of 640 kpc from the cluster center and has a 50 kpc long collimated \ha tail that stretches directly north of the galaxy \citep{gavazzi01}. In Figure \ref{97073img}, we show images at IRAC 3.6 and 8 $\mu$m, \ha (obtained from the GOLDMine database), and rest 17.035 $\mu$m (H$_2$ 0-0 S(1) transition) wavelengths. Moreover, we show the approximate location and orientation of the faint \ha tail, seen by \cite{gavazzi01} in their deep H$\alpha$ images of this galaxy, with an arrow. The IRS slit positioning captures only a small portion of the \ha tail. The side of the galaxy thought to be interacting with the ram-pressure wind is the southernmost edge. From a comparison of the H$\alpha$, 8 $\mu$m, and 17 \mic images, it is clear that the 17 \mic emission does not fully match the observed \ha and 8 \mic emission even after considering the effects of spatial resolution. The overabundance of 17 \mic emission at the tail end of the galaxy suggests the presence of warm H$_2.$

\begin{deluxetable*}{ccccccc}
\tablewidth{0pt}
\small
\centering
\tablecolumns{6}
\tablecaption{Measured H$_2$ Properties and Gas Masses for Galaxies \label{gasmass}}
\tablehead{
\colhead{Galaxy} &
\colhead{Region} & 
\colhead{Component} &
\colhead{T$_{\rm ex}$ (K)} &
\colhead{$N_{\rm tot}$ (cm$^{-2}$)} &
\colhead{$\Sigma$ (M$_\odot$ pc$^{-2}$)} &
\colhead{H$_2$ Mass (M$_\odot$)}
}
\startdata
CGCG 97073 	& Full & Warm &$130^{+6}_{-12}$ & $6.6^{+2.0}_{-0.8}\times10^{19}$ & $1.1^{+0.3}_{-0.1}$ & $4.7^{+1.4}_{-0.5}\times10^8$ \\
			&  & Hot & $593^{+233}_{-115}$ & $2.6^{+3.1}_{-1.5}\times10^{17}$ & $4.2^{+4.9}_{-2.4}\times10^{-3}$ &  $1.8^{+2.2}_{-1.0}\times10^6$\\
			& Tail & Warm &$114^{+17}_{-13}$ & $1.2^{+0.6}_{-0.4}\times10^{20}$ & $1.9^{+0.9}_{-0.6}$ & $2.1^{+1.0}_{-0.7}\times10^8$ \\
			& & Hot & $480^{+314}_{-44}$ & $6.5^{+3.4}_{-5.0}\times10^{17}$ &  $1.0^{+0.5}_{-0.8}\times10^{-2}$ &   $1.2^{+0.6}_{-0.9}\times10^6$\\
\tableline
NGC 4522 	& Central & Warm & $116^{+13}_{-14}$ & $1.9^{+1.0}_{-0.5}\times10^{20}$ & $3.0^{+1.6}_{-0.8}$ & $3.7^{+2.1}_{-1.0}\times10^7$ \\
			& & Hot & $423^{+123}_{-42}$ & $1.3^{+0.8}_{-0.9}\times10^{18}$  & $2.1^{+1.3}_{-1.4}\times10^{-2}$ & $2.7^{+1.6}_{-1.7}\times10^5$ \\
		 	& NE & Warm & $141\pm2$ & $4.0\pm0.3\times10^{19}$ & $0.64\pm0.04$ &  $4.8\pm0.3\times10^6$ \\
		 	& SW & Warm & $127\pm2$ & $6.4^{+0.6}_{-0.5}\times10^{19}$ & $1.0\pm0.1$ & $8.7^{+0.8}_{-0.7}\times10^6$  \\
\tableline
NGC 1427A 	&  Galactic & Warm & $< 91$ &  $>1.2\times10^{20}$ & $>1.8$ & $>5.9\times10^7$ \\
			&		 & Warm\tablenotemark{a} & $< 104$ &  $>6.6\times10^{19}$ & $>1.1$ &  $>3.4\times10^7$ \\ 	
			&  High IR Flux & Warm & $<91$ &  $>1.7\times10^{20}$ & $>2.8$ & $>4.1\times10^{7}$ \\
			&		 & Warm\tablenotemark{a} & $< 103$ &  $>1.0\times10^{20}$ & $>1.6$ & $>2.4\times10^{7}$
\enddata
\tablenotetext{a}{This assumes an OPR of 1.5.}
\end{deluxetable*}

\par
\begin{figure*}
\epsscale{0.75}
\plotone{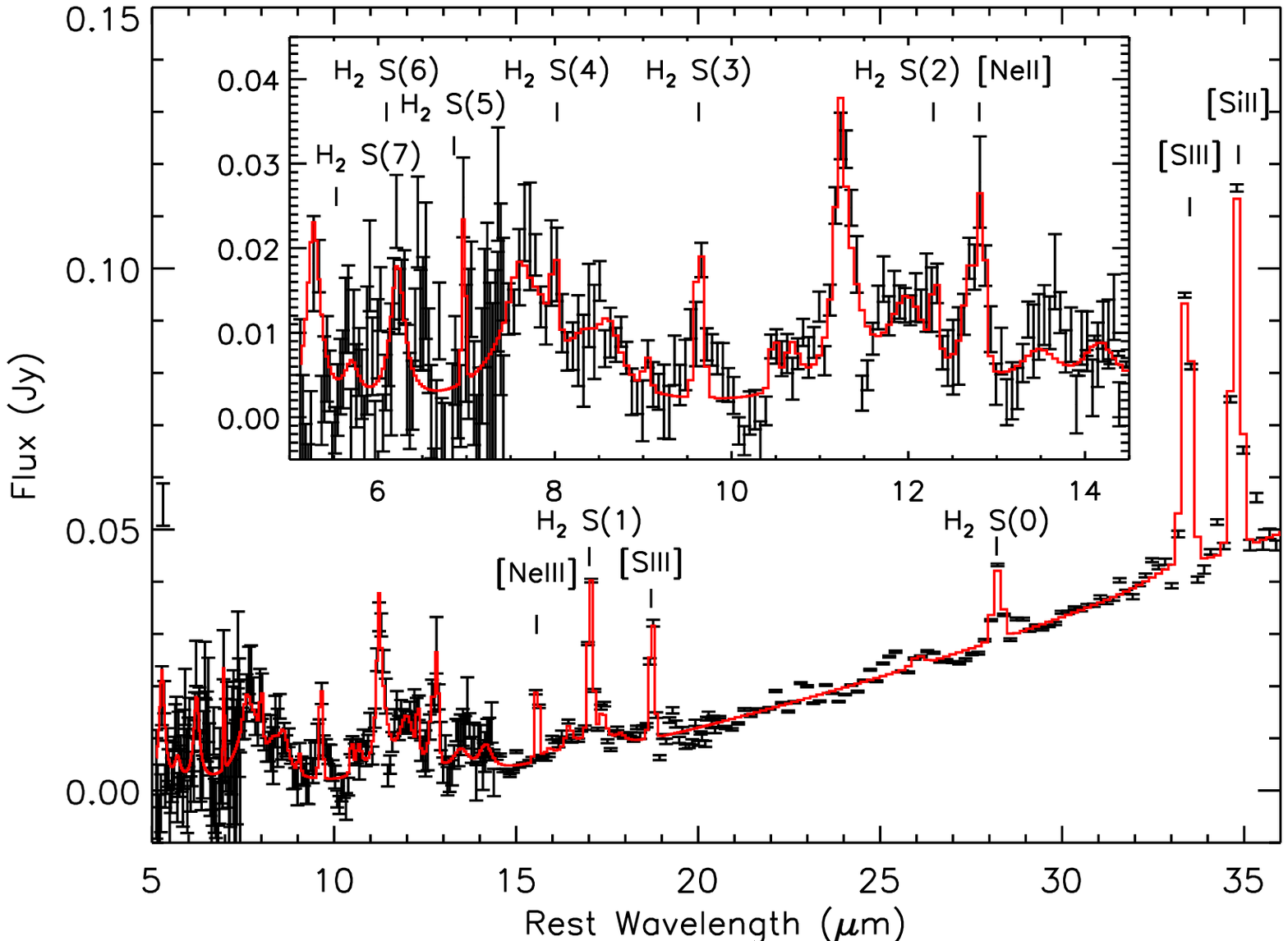}
\plotone{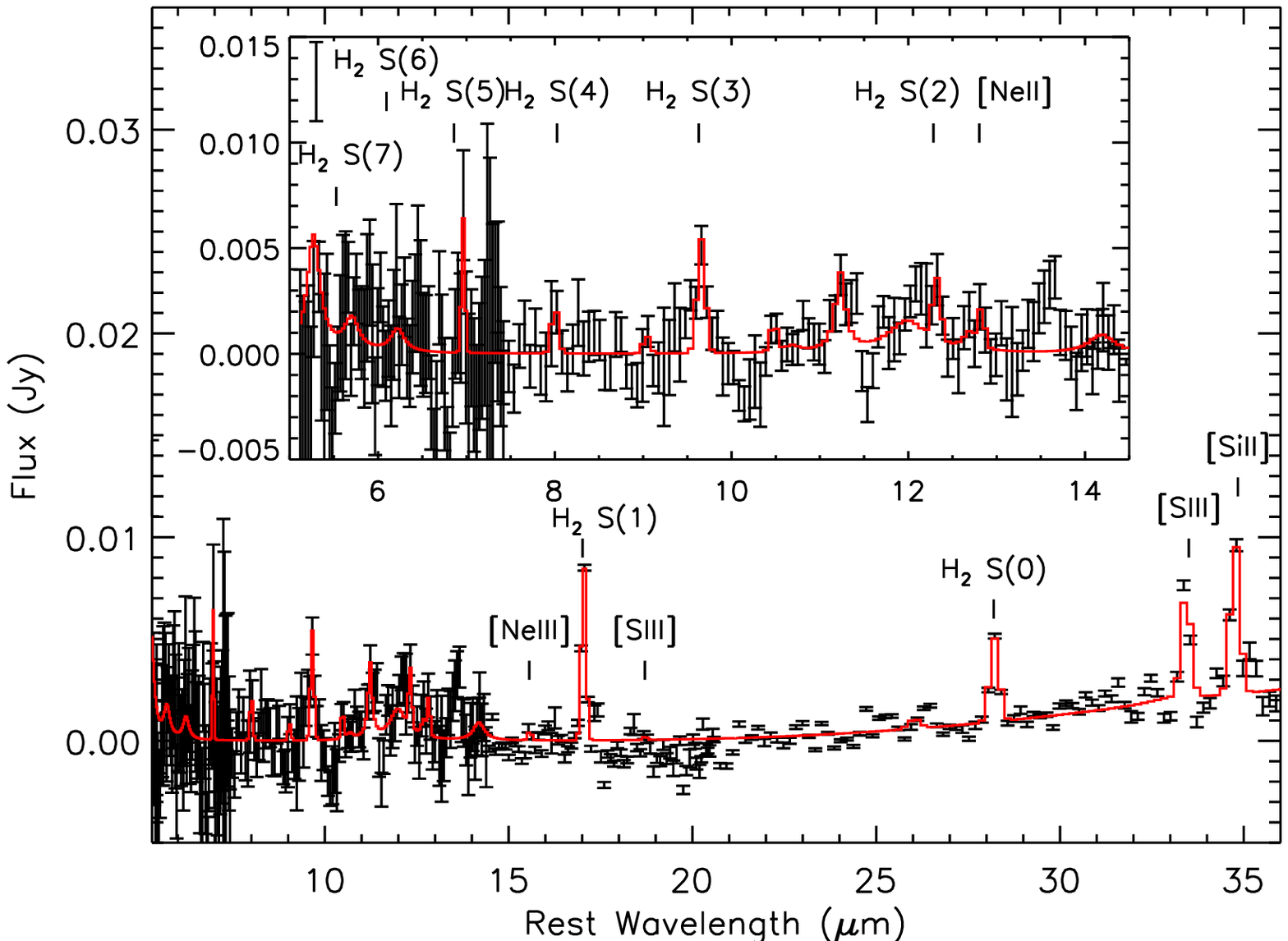}
\caption{Infrared spectra of 97073 extracted from the full galaxy extraction region (top) and tail extraction region (bottom). Both spectra show significant detections of ground-state rotational H$_2$ lines. In both regions, the S(0) thru S(3) lines are detected. In the tail extraction region, there is significant line emission but very little continuum emission, which suggests that star formation is not the primary excitation source for H$_2$ in this region. \label{97073spec}}
\end{figure*}

Upon closer inspection, we detect significant emission of warm \htwo within the galaxy and in a portion of its tail in the extracted spectra. We show the extracted spectra within the full galaxy extraction region and tail extraction region in Figure \ref{97073spec}. Within the full galaxy extraction region, \htwo lines in addition to a few fine structure lines dominate the spectrum. There are significant detections of the ground state rotational H$_2$ S(0) to S(3) lines. In the tail extraction region, the strongest lines are \htwo lines with the S(1) line being the strongest. What is particularly striking within this region is the large ratio of the \htwo line relative to continuum emission. It is clear that a significant fraction of the flux in the spectrum arises from \htwo line emission from the tail region.  The \htwo emission of this galaxy is as dramatic as was observed in ESO 137-001, which is discussed at length in Paper 1. Bright fine structure lines, such as [NeII], [NeIII], [SIII], and [SiII], are also observed in both extraction regions.

\par
\begin{figure}
\centering
\epsscale{1.2}
\plotone{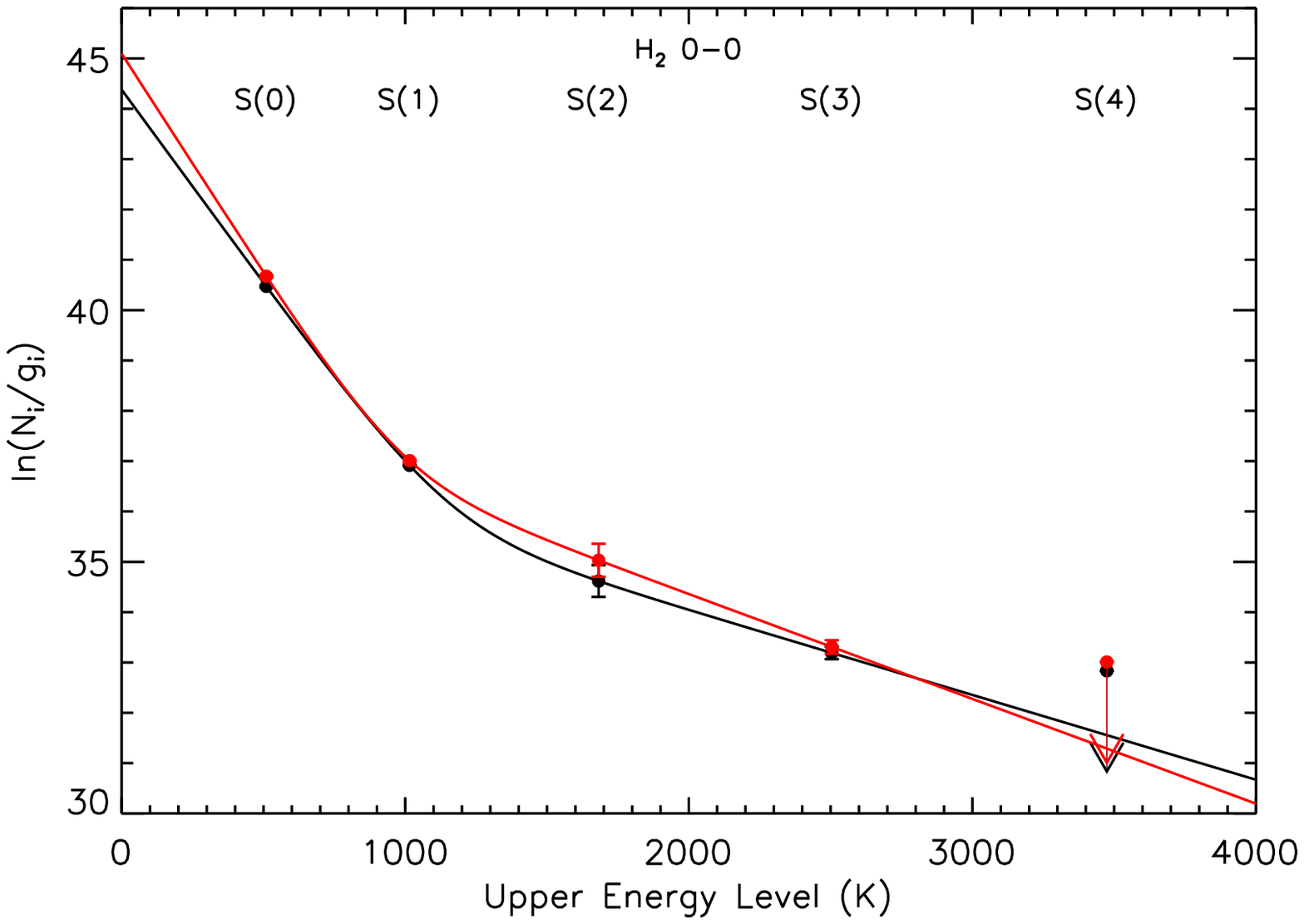}
\caption{H$_2$ excitation diagram of 97073 for both the full galaxy extraction region (black points) and the tail extraction region (red points). The upper limits for non-detections shown are 3$\sigma.$ The black and red curves are fit to the data in the full galaxy and tail regions, respectively. We fit the data with a two-temperature H$_2$ model and obtain a satisfactory fit for the observed line detections S(0) thru S(3) in both cases. The excitation of the gas in both regions is largely similar. \label{97073excitation}}
\end{figure}

We calculate the column density, temperature, and total mass of the \htwo gas in the two regions using the methods outlined in Paper 1.  To calculate the column density associated with each line transition, we use solid angles, $\Omega$, of $5.16\times10^{-8}$ and $1.30\times10^{-8}$ sr for the galaxy and tail extraction regions, respectively. The excitation diagrams of the H$_2$ ground-state rotational transitions for the two regions are shown in Figure \ref{97073excitation}. The column densities for both regions cannot be fit with a single temperature model as there appear to be multiple temperature components. This was also the case for ESO 137-001. We adopted the same two-temperature-component model discussed in Paper 1. The model (a hot and a warm component; Equation 4 in Paper 1) fits adequately for both extraction regions. The fit results are tabulated in Table \ref{gasmass}. For the whole galaxy, we determine the warm component to have a temperature of $130^{+6}_{-12}$K and a column density of $6.6^{+2.0}_{-0.8}\times10^{19}$ cm$^{-2}.$ For the hot component, we find a temperature of $593^{+233}_{-115}$K and a column density of $2.6^{+3.1}_{-1.5}\times10^{17}$ cm$^{-2}.$ 

\par
The errors were estimated using Monte Carlo simulations. Because there were no additional degrees of freedom in the fit given that there are an equal number of data points as there are fit parameters, a new set of best fit model parameters was derived from each trial run. This assumed that each flux data point was Gaussian distributed with a standard deviation that was equal to its 1$\sigma$ error and that the data can only be fit by a two temperature model. On the occasion where the fit failed to converge for a given trial, i.e. the two temperature model was not a good fit, that trial run was discarded. This occurs when the $\ln(N/g)$ data points are not monotonically decreasing as a function of transition temperature, which is most likely unphysical. We carried out 10,000 successful trials to determine the distribution of 1$\sigma$ errors in the model parameters. The number of trials was sufficiently large to ensure the convergence of the error estimates. For all error estimations in this paper, we follow the same method. 
\par
For the tail region, we derived temperatures for the warm and hot component of $114^{+17}_{-13}$K and $480^{+314}_{-44}$K, respectively. These components had corresponding column densities of $1.2^{+0.6}_{-0.4}\times10^{20}$ cm$^{-2}$ and $6.5^{+3.4}_{-5.0}\times10^{17}$ cm$^{-2}.$ The values of these fits are consistent with those derived for the full galaxy. In both cases, the warm component mass dominates over that of the hot component by a significant margin. The column densities and temperatures for \htwo in these two regions are also similar to the values obtained for the tail in ESO 137-001, suggesting that there is similar physics exciting the \htwo gas in both galaxies. 
\par
We measure a total warm H$_2$ mass of $4.7^{+1.4}_{-0.5}\times10^8$ M$_\odot$ and $2.1^{+1.0}_{-0.7}\times10^8$ M$_\odot$ within the full galaxy and tail regions, respectively. This indicates that a significant fraction of the total galaxy warm H$_2$ is found within the tail region. We compare the warm \htwo mass measurement with other estimates of gas masses measured for this galaxy such as cold \htwo and HI masses. In particular, we compare the warm-to-cold \htwo to look for any peculiarities. We determine the cold \htwo mass of this galaxy by using the \htwo measurement obtained from \cite{boselli94} who carried out CO observations of the galaxy; we applied the appropriate distance and $X_{CO}$ corrections to account for changes in their assumed values. \cite{boselli94} assumed a distance of 65 Mpc to this galaxy and used an $X_{CO}$ value of $2.3\times10^{20}\:\textrm{cm}^{-2}/\textrm{K-km s}^{-1}.$ For our work, we adopt an $X_{CO}$ of $2.8\times10^{20}\:\textrm{cm}^{-2}/\textrm{K-km s}^{-1}$ from \cite{bloemen86}, giving an \htwo mass of $1.7\times10^9$ M$_\odot.$ This yields a warm-to-total \htwo mass fraction of 0.21. We assume that the aperture correction effects between the CO measurements and our \htwo measurements are not significant. This fraction is higher than the highest mass fraction value reported by \cite{roussel07} in their warm H$_2$ survey of {\it Spitzer} SINGS galaxies with published CO measurements. This is also a factor of two higher than the fraction of 0.1 reported by \cite{rigopoulou02} for starburst galaxies in their warm \htwo survey. There has been some suggestion in Stephan's Quintet that shock-heated \htwo may have an anomalously high fraction of warm-to-cold \htwo gas \citep{guillard12}, which also may be the case here. We also compare the warm \htwo-to-total gas fraction of the galaxy. We adopt the HI gas mass of this galaxy of $2.0\times10^9$ M$_\odot$ measured by \cite{scott10}. We obtain a fraction of 0.11, which is a significant fraction of the total gas within the galaxy.
\par
\begin{figure}
\epsscale{0.8}
\plotone{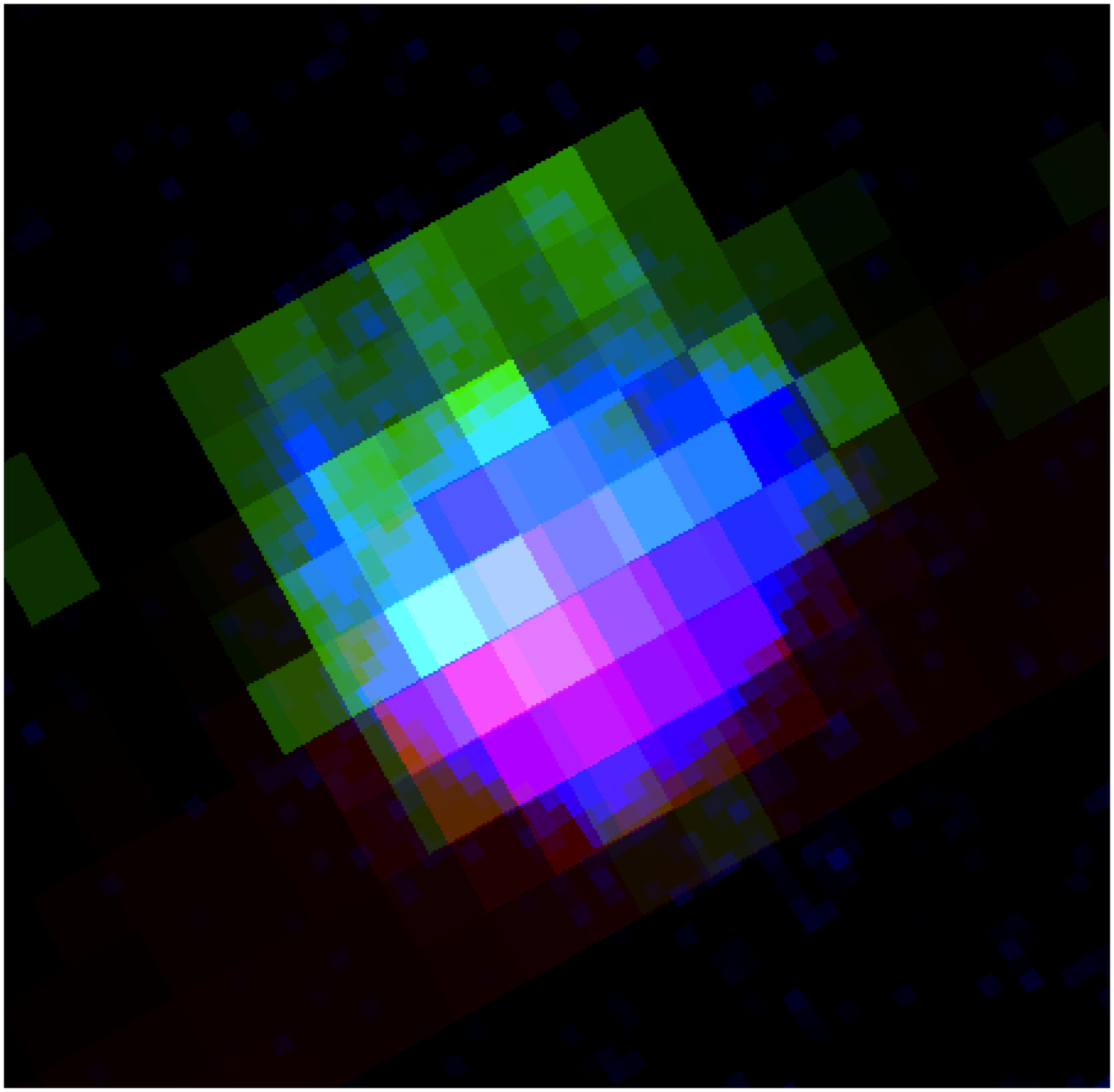}
\caption{This is a color composite of 97073 seen at three difference wavelengths. The blue image is the 8 $\mu$m excess image, the green is the continuum-subtracted 17 $\mu$m H$_2$ image, and finally the red is the 24 $\mu$m image. There is clear offset in the H$_2$ emission compared to the 24 $\mu$m emission, which is associated with dust in star-forming regions. \label{97073mips}}
\end{figure}

We test to see if star-formation is the source of the H$_2$ excitation by comparing the 24 $\mu$m image with the continuum-subtracted H$_2$ image.  Figure \ref{97073mips} presents our results. The 24 $\mu$m emission (red) is mainly confined to the part of the galaxy directly opposing the tail, presumably the location where the ICM wind hits the galaxy. The emission appears to be mainly confined to the location where most of the star formation is occurring. The blue image is the 8 $\mu$m aromatic emission; the green image represents the continuum-subtracted H$_2$ S(1) transition image. The H$_2$ emission is clearly offset from the 24 $\mu$m emission toward the direction of the tail. 

\subsubsection{NGC 4522}
\begin{figure*}
\epsscale{0.75}
\plotone{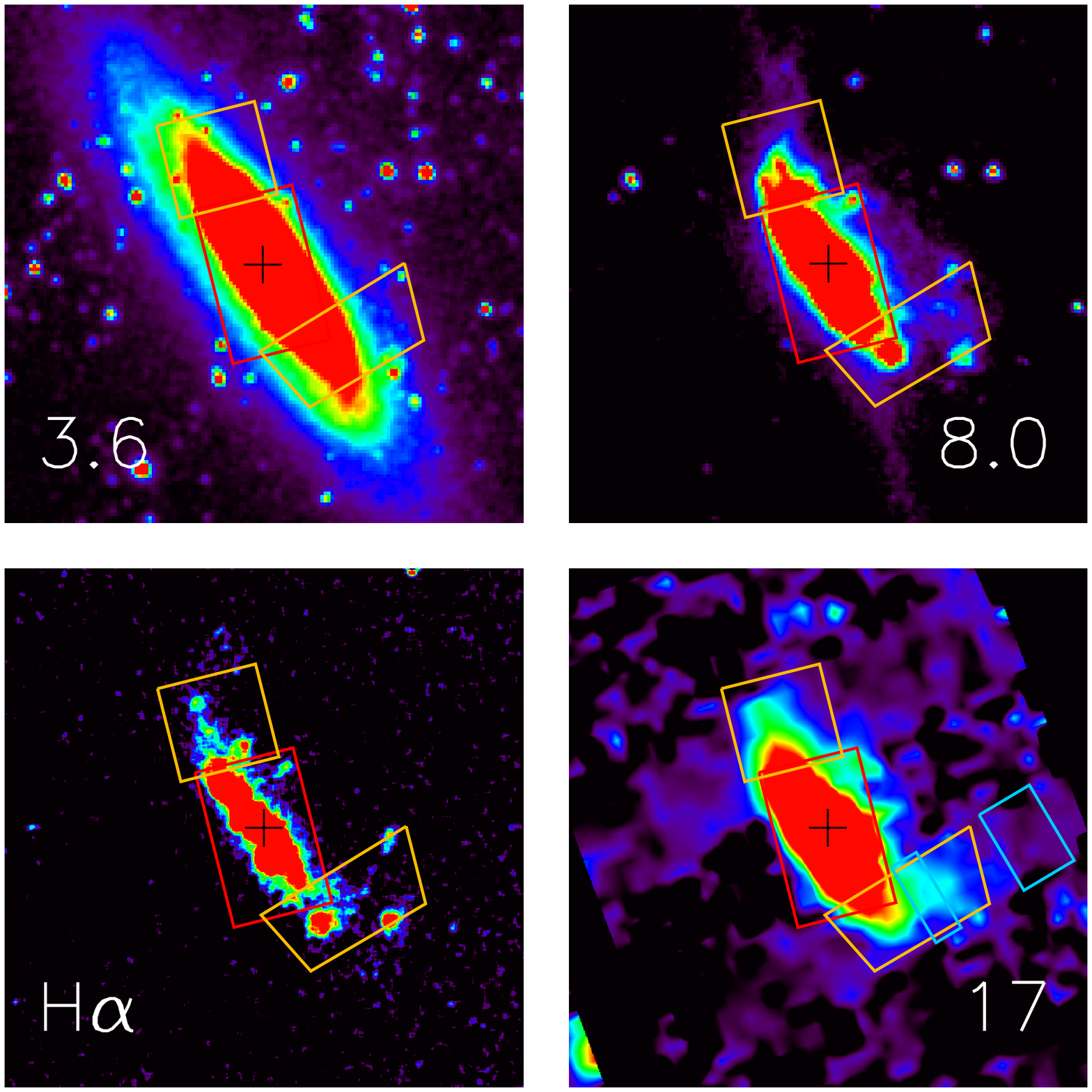}
\caption{\footnotesize 2$\arcmin\times$2$\arcmin$ images of NGC 4522 at 3.6 $\mu$m, 8 $\mu$m, H$\alpha$ (obtained from GOLDMine), and rest 17.035 $\mu$m (H$_2$ 0-0 S(1) transition) wavelengths. North is up and East is left. The colors follow the visible spectrum where blue represents the faintest emission and red the brightest. The center of the galaxy is shown by the black cross. The red, orange, and blue boxes represent spectral extraction regions. The red box is the central extraction region, the orange ones the NE and SW regions where ram-pressure stripping is occurring, and finally the blue extraction region is used to study the kinematics of the stripped gas. Even though at 3.6 $\mu$m this galaxy looks like a typical spiral, there are strong asymmetries observed at H$\alpha$, 8, and 17 $\mu$m. Warm dust is being blown out of the galaxy as seen by the dust trails at 8 $\mu$m. The 17$\mu$m image was generated by coadding both our LL and archival LL observations. It is dominated by both the 17 $\mu$m aromatic feature and the H$_2$ line, so it is difficult to ascertain if there are locations with anomalously high H$_2$ emission. However, a tail-like feature is observed within the SW orange extraction region. \label{n4522img}}
\end{figure*}

NGC 4522 is an infalling spiral located within the Virgo cluster and is at a projected separation of 1.1 Mpc from the cluster center. This galaxy has a large radial velocity of 1250 km s$^{-1}$ with respect to the cluster (obtained from NED), which suggests that a large portion of its motion is along our line-of-sight. We adopt a distance to Virgo of 16.5 Mpc \citep{mei07}. We present the {\it Spitzer} IRAC, IRS, and GOLDMine \ha images of NGC 4522 in Figure \ref{n4522img}. The 3.6 $\mu$m image of the galaxy shows a typical unperturbed spiral. However, the \ha, 8, and rest 17 \mic images reveal something completely different. Their emission is truncated compared to the radial extent of the stellar disk of the galaxy. We also see extraplanar tail-like features in the southwest quadrant of the galaxy at 8 \mic and 17 $\mu$m, which is a clear signature of dust and possibly gas stripping. The undisturbed nature of the stellar disk indicates that ram-pressure and not tidal forces must be at play. However, this explanation is somewhat complicated by the fact that the galaxy is approximately 1 Mpc away from the Virgo cluster center and the traditional view is that the galaxy velocity and the density of the ICM at that distance cannot be sufficient to produce significant ram-pressure. However, it is likely that the relative velocity of NGC 4522 within the ICM is much higher than expected either because the ICM is dynamic, i.e. it is moving relative to the cluster mean velocity, and/or the galaxy is traveling at high speeds and is unbound from the cluster \citep{vollmer06}. 

\begin{figure*}
\epsscale{0.75}
\plotone{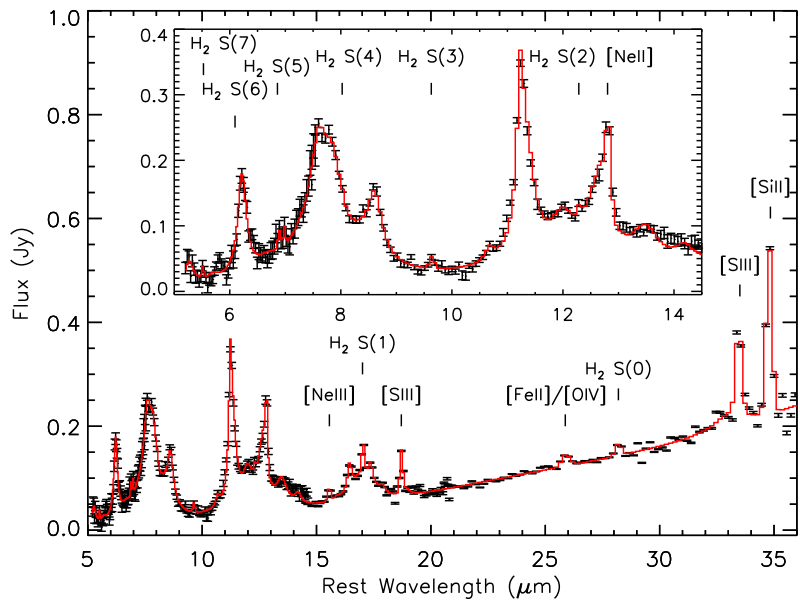}

\caption{Infrared spectrum of NGC 4522 extracted from the central extraction region (shown as the red box in Figure \ref{n4522img}). The spectrum shows significant detections of ground-state rotational H$_2$ lines. The S(0) thru S(3) lines are detected.  \label{n4522spec} }
\end{figure*}

\begin{figure*}
\epsscale{0.75}
\plotone{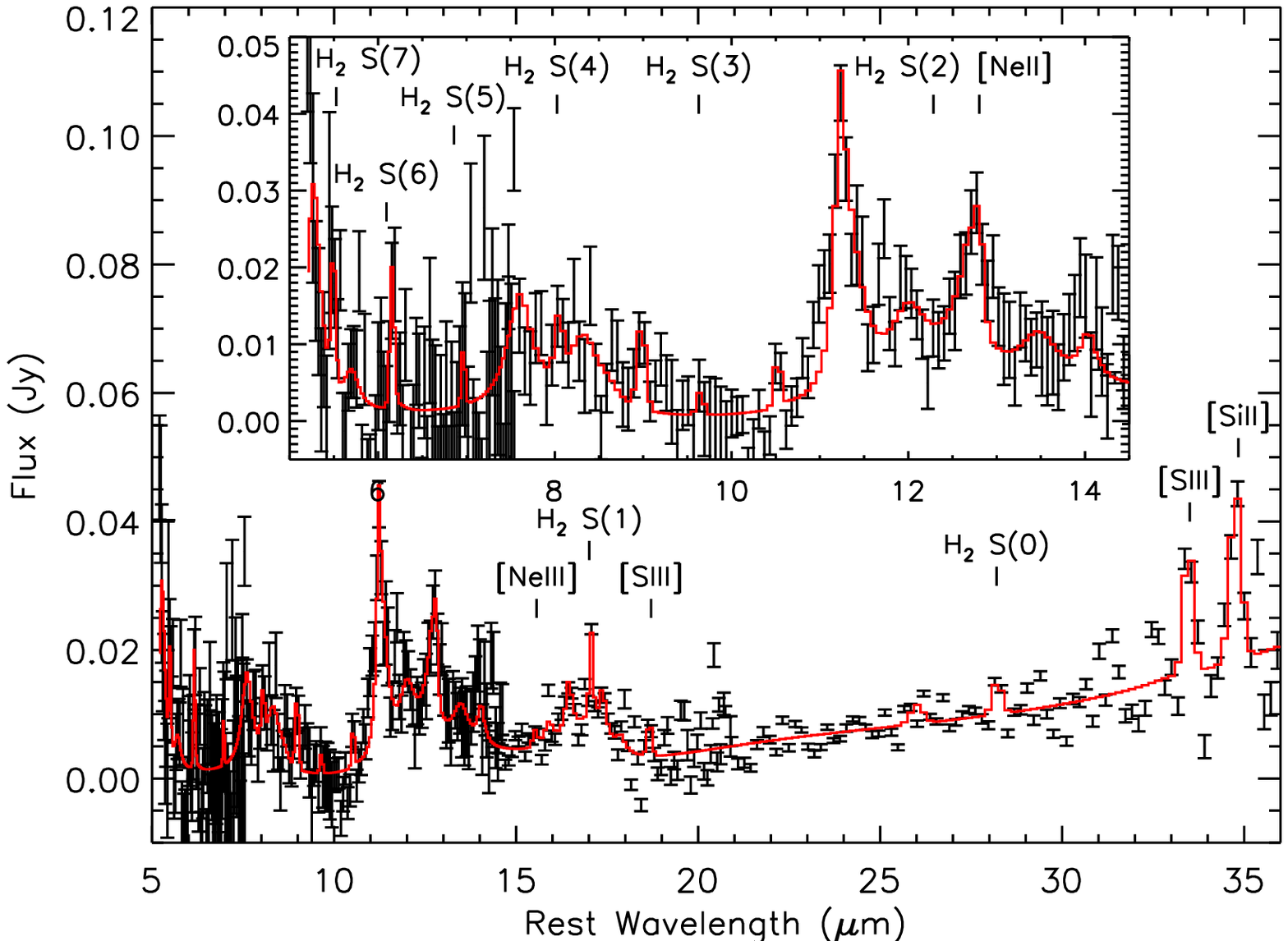}
\plotone{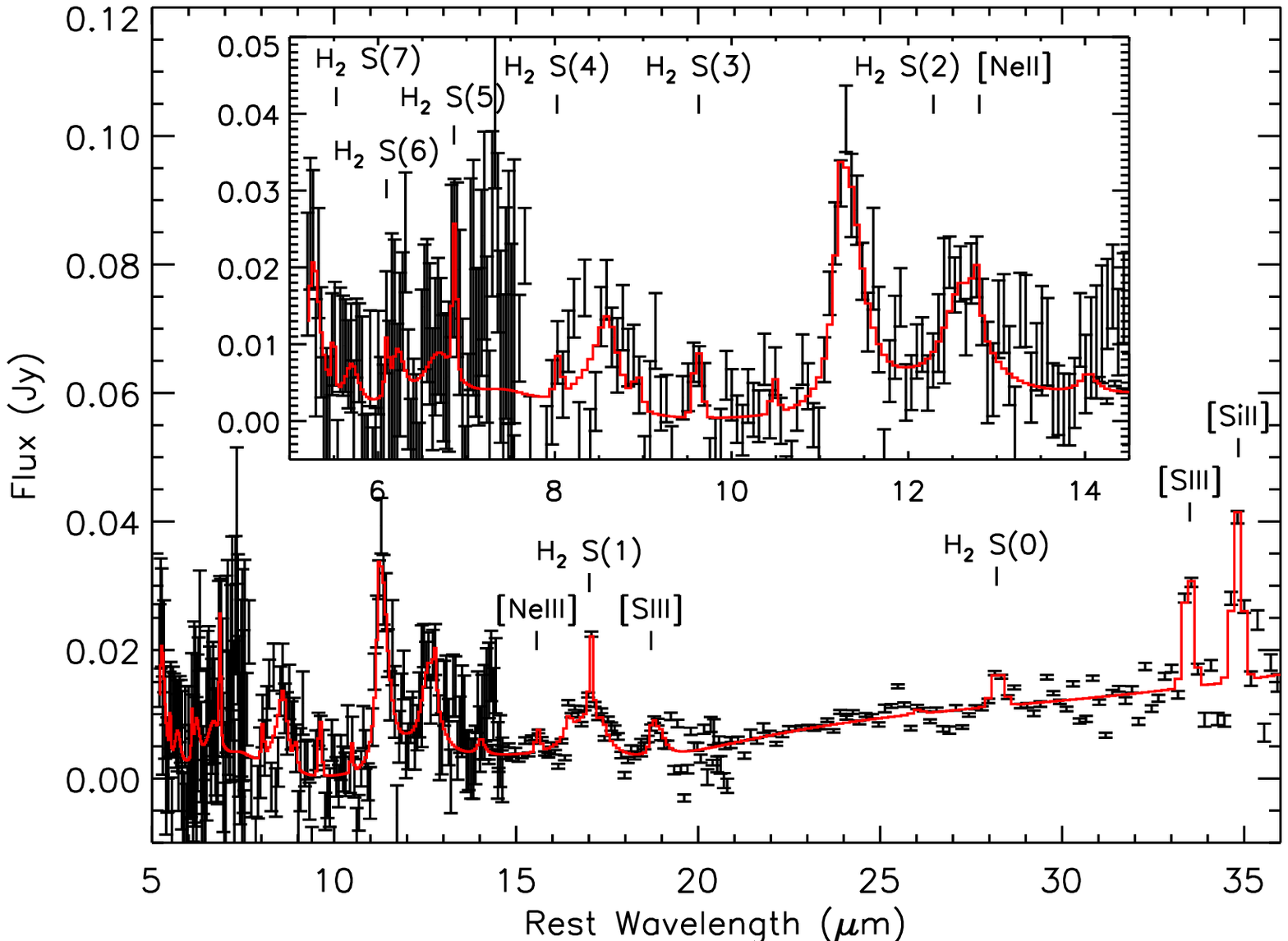}
\caption{Infrared spectra of the NE and SW regions in NGC 4522 (shown as orange boxes in Figure \ref{n4522img}) where ram-pressure is thought to be stripping gas. The top plot shows the spectrum extracted from the NE extraction region while the bottom plot shows the spectrum extracted from the SW part of the galaxy. Significant detections of the \htwo S(0) and S(1) lines are observed. \label{n4522tail}}
\end{figure*}
\par
We carry out a more careful analysis by extracting spectra from key regions within the galaxy, as shown in Figure \ref{n4522img}. In Figure \ref{n4522spec}, we present the spectra extracted from the central extraction region shown in Figure \ref{n4522img}. We detect warm \htwo in a tail-like feature that extends approximately 4 kpc in length from the midplane of the disk along the SW extraction region. The central extraction region was defined to obtain most of the infrared emission originating from the galaxy. In it, we have significant detections of the ground-rotational level \htwo S(0) thru S(3) lines. We also observe multiple fine structure lines and significant emission by dust, likely associated with star formation. We will explore the meaning of this observation in greater detail in Section \ref{starformation}. We also extract spectra from the northeast (NE) and southwest (SW) regions of the 17 \mic emission (see Figure \ref{n4522img} for the exact location of the extraction regions) where the effects of ram-pressure are most obvious. The SW region spectrum was extracted from the higher signal-to-noise archival dataset. Due to the poorly constrained and unphysical value of the extinction parameter obtained by \emph{PAHFIT}, we fixed the optical depth of the 9.7\mic silicate absorption feature to that measured for the central extraction region for the fits of the SW and NE regions. In both regions, we detect both the \htwo S(0) and S(1) lines. The SW region has a detection of the S(7) line while the NE region has a detection of the S(3) line. The S(7) line detection is not particularly convincing because of the increasing noise in that region. We do not include it in our model fits. It is clear that the SW region, which contains most of the observed extraplanar 17 \mic emission, contains a significant amount of warm H$_2.$ We also observe aromatic features, fine-structure lines, and dust continuum, though the relative ratio of the \htwo and dust emission appears smaller when compared to the central extraction region. We investigate the implications of this difference in further detail in Section \ref{singscomparison}. 
\begin{figure}
\epsscale{1.2}
\plotone{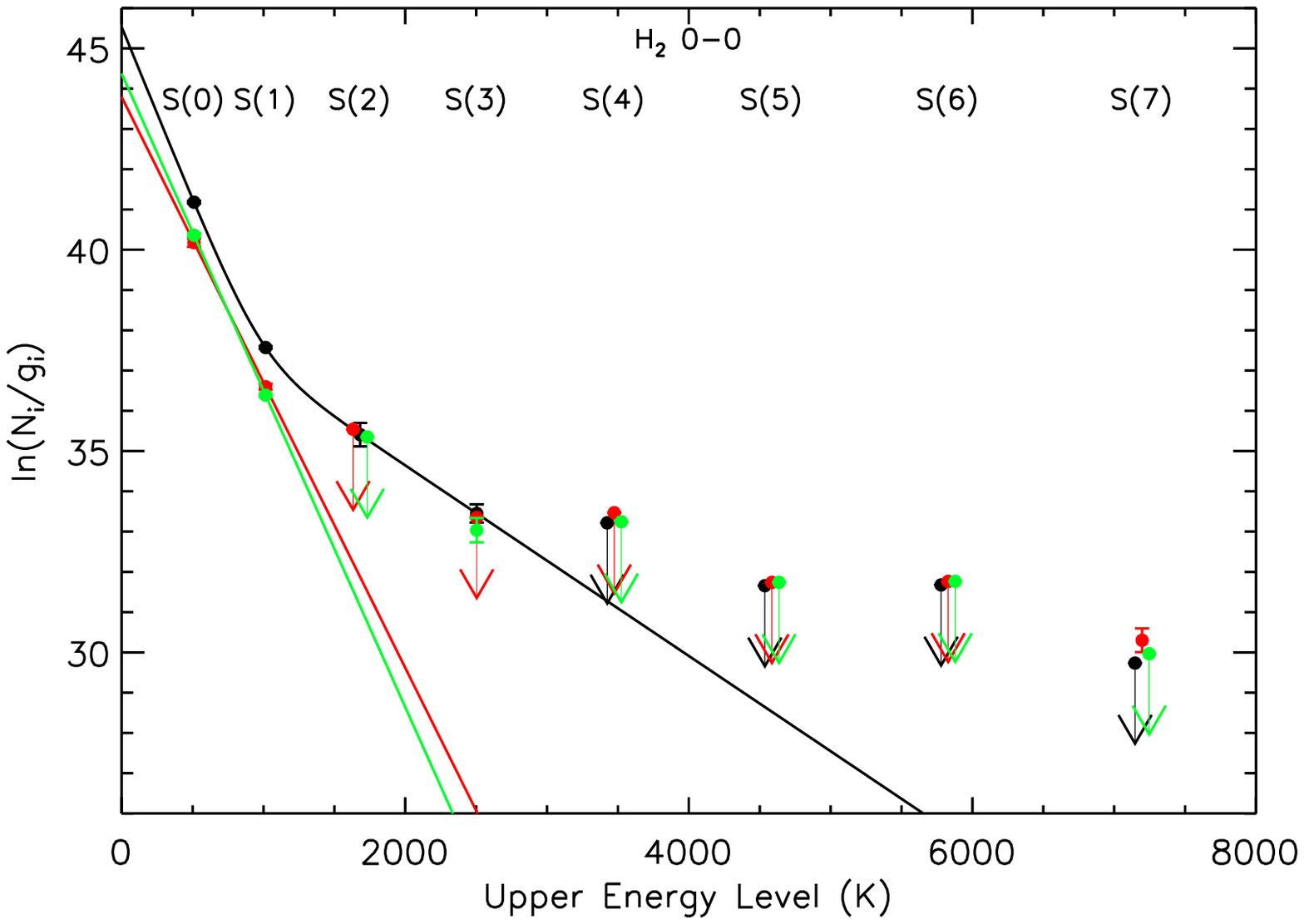}
\caption{H$_2$ excitation diagram for NGC 4522 for three different extraction regions. The black points represent the H$_2$ in the central region of the galaxy. The red points are for the NE region, and the green ones are for the SW region. The upper limits for non-detections shown are 3$\sigma$ limits. The limits for the three different regions are offset from each other to improve readability. We fit a two temperature component model for the data from the central region to account for the changing slope in the excitation diagram. The best fit model is shown by the black curve. For the NE and SW region, we fit a one temperature model because we only have firm detections of the S(0) and S(1) lines. The NE and SW regions do have a detection of a higher excitation line, but at least four lines are required to fit a two temperature model. The best fit models for the NE and SW regions are shown by the red and green lines, respectively. \label{n4522excitation}}
\end{figure}

\par
To understand the physical conditions of the \htwo present in the galaxy and its tail, we fit an excitation model to the H$_2$ line fluxes obtained from the three different spectral extraction regions using the same methods used for 97073. The solid angles for the central, NE, and SW regions are $4.58\times10^{-8},$ $2.75\times10^{-8},$ and $3.10\times10^{-8}$ sr, respectively. The results of these fits are also given in Table \ref{gasmass}. For the central region, we fit a two-temperature model, which yields a warm component with a temperature of $116^{+13}_{-14}$K and column density of $1.9^{+1.0}_{-0.5}\times10^{20}$ cm$^{-2}$ and a hot component with a temperature of $423^{+123}_{-42}$K and a column density of $1.3^{+0.8}_{-0.9}\times10^{18}$ cm$^{-2}$. For the NE region, we only fit a single temperature model as we only have significant detections of two lines. We obtain a temperature of $141\pm2$K and a column density of $4.0\pm0.3\times10^{19}$ cm$^{-2}$. For the SW region, we also fit a single temperature model due to only three significant detections of \htwo lines. We fit the S(0) and S(1) lines in this case. We obtain a temperature of $127\pm2$K and a column density of $6.4^{+0.6}_{-0.5}\times10^{19}$ cm$^{-2}$. The warm components of all three regions have largely consistent temperatures. We calculate a total warm H$_2$ mass of $3.7^{+2.1}_{-1.0}\times10^7$ M$_\odot,$ $4.8\pm0.3\times10^6$ M$_\odot,$ and $8.7^{+0.8}_{-0.7}\times10^6$ M$_\odot$ for the central, NE, and SW regions, respectively. This is an order of magnitude less warm H$_2$ mass than was measured for 97073. We also compare the central region warm \htwo value with the measured cold gas mass values to see if the warm-to-cold gas ratio is abnormal in any way. This galaxy has a HI mass of $3.8\times10^8$ M$_\odot$ \citep{kenney04}, and a cold H$_2$ mass of $3.2\times10^8$M$_\odot,$ derived using CO measurements \citep{smith97} corrected to our adopted distance to the galaxy and $X_{CO}.$ This corresponds to a warm-to-cold H$_2$ fraction of 0.15 and a warm-to-total gas fraction of 0.06. This is largely consistent with what is observed in the SINGS sample with warm and cold H$_2$ measurements \citep{roussel07}.
\par
\begin{figure}
\epsscale{1.2}
\plotone{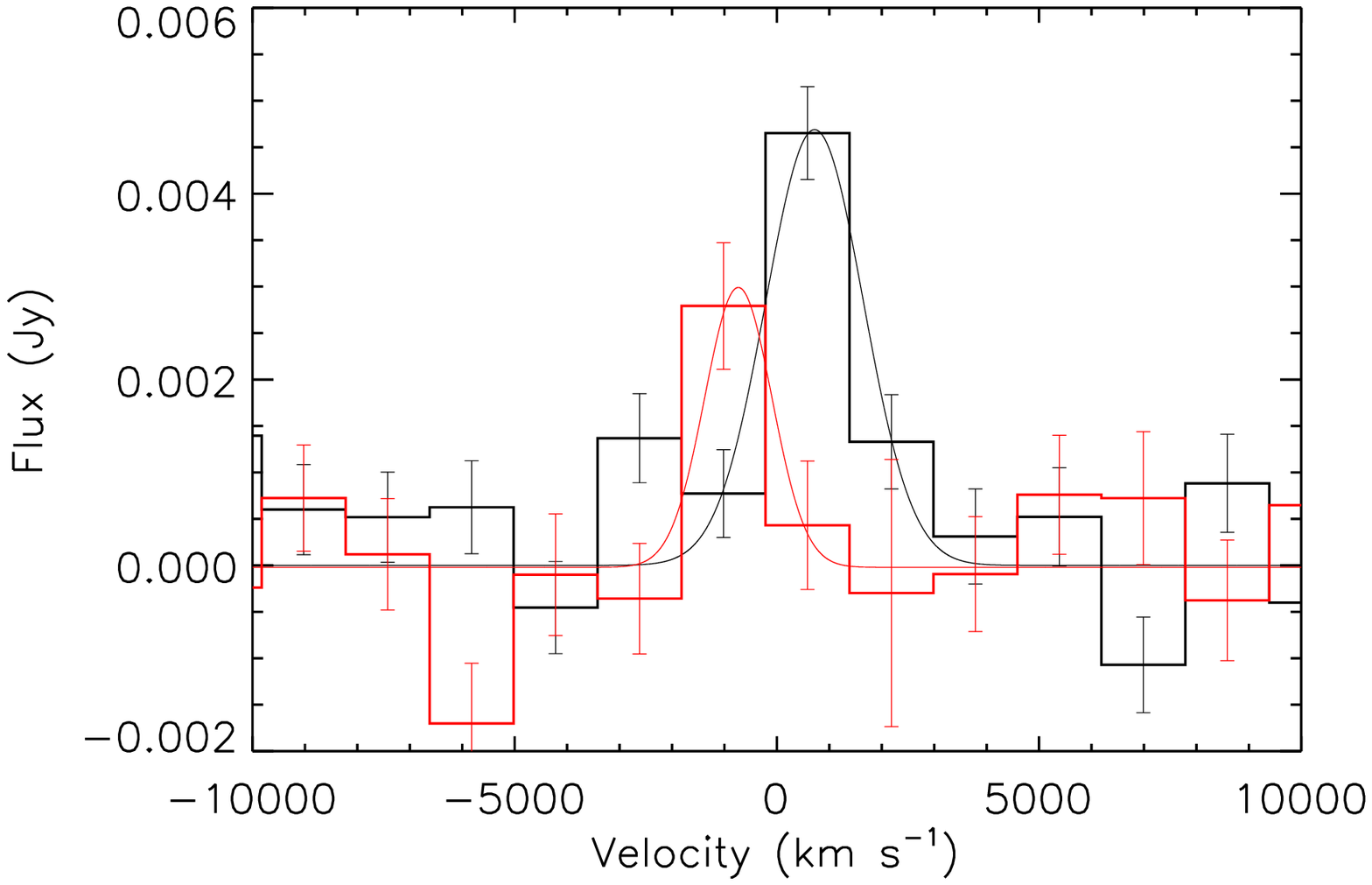}
\caption{Continuum-subtracted line profiles of the H$_2$ S(1) line at two different locations in the stripped tail of NGC 4522. The galaxy's line-of-sight velocity is set to be zero. An object with negative velocity means that it is moving away from the galaxy toward us. The black histogram is the line profile for a region of the tail closest to the galaxy whereas the red histogram is the line profile for a location in the tail that is further away from the galaxy. The solid curves represent the respective gaussian line profile fits. There is a clear velocity shift in the \htwo emission as one moves further away from the galaxy. The gas further away from the galaxy has a lower line-of-sight velocity. The velocity offset is consistent in direction and magnitude to the velocity difference between the galaxy and the cluster mean. This means that the gas may be being stripped permanently from the galaxy. \label{n4522offset}}
\end{figure}

We looked for kinematic signatures of the warm \htwo gas being stripped permanently from the galaxy. The galaxy has a high enough recessional velocity with respect to the cluster average that, if the stripped gas were to be permanently lost to the cluster, then the velocity shift of a spectral line should be observable by the IRS low resolution spectrographs. We chose the 17.035 \mic S(1) line for this measurement. We use two different extraction regions along the \htwo tail detected in the SW region: one close to the galactic disk and one farthest away along the tail where a reliable line detection can be obtained. The extraction regions are shown in the 17 \mic image in Figure \ref{n4522img} as cyan boxes. The continuum-subtracted line profiles from each of these regions are shown in Figure \ref{n4522offset}. There is a clear blueward shift of the S(1) line as one moves from just outside of the disk of the galaxy to the farthest reaches of the detected tail. 

\par
We carry out a gaussian fit to the line profile of each extraction region to quantify the shift in velocity. The fit velocity and FWHM parameters are $v = 720\pm170$ and $\Delta v = 2190\pm280$ km s$^{-1}$ for the region close to the disk. For region along the tail, we obtain fit values of $v = -740\pm260$ and $\Delta v = 1510\pm490$ km s$^{-1}.$ This gives a relative velocity difference of $\sim 1460\pm310$ km s$^{-1}.$ The relative shift in velocity is in the right direction and magnitude if the gas is lost permanently from the cluster because the measured relative velocity of the galaxy with respect to the cluster mean is 1250 km s$^{-1}.$ To further check the validity of the shift, we investigate the properties of the IRS spectrograph used for these detections. The spectral resolutions of the low resolution spectrographs are not fixed and vary with wavelength. The spectral resolution at the 17 \mic line with the LL2 spectrograph is approximately 100. This corresponds to a velocity width of an unresolved line of $\sim3000$ km s$^{-1},$ with two pixels per resolution element. With these spectrograph characteristics and our achieved signal-to-noise one should be able to detect a shift of approximately a quarter of a line width, which corresponds to a velocity of 750 km s$^{-1}.$ This value is slightly larger than the wavelength uncertainty in the calibration of 600 km s$^{-1}.$ We conclude that we may have detected a velocity shift. No other mid-IR line, such as \htwo S(0) or [SiII] 34.8 $\mu$m, could be identified in the extraction region farthest from the galaxy to confirm this result. If this tentative result is correct, some of the stripped gas is being permanently lost to the ICM. Other measurements that might confirm this interesting result are inconclusive. \cite{kenney04} also see a blueward shift in their Very Large Array (VLA) HI measurements of the stripped extraplanar gas in this galaxy. However, the blueshift is only modest, at the level of $10-20$ km s$^{-1},$ and does not extend as far from the galactic plane as our measurement. CO observations were carried out by \cite{vollmer08}, but they are even less extended than the HI map. 

\subsubsection{NGC 1427A}

\begin{figure*}
\epsscale{0.75}
\plotone{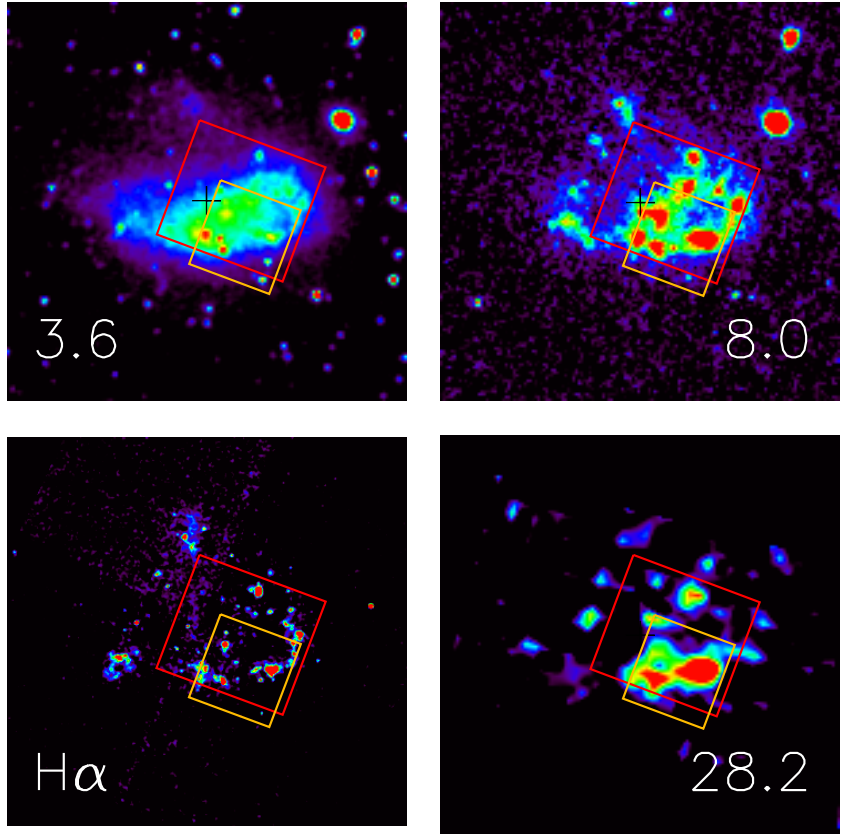}
\caption{2$\arcmin\times$2$\arcmin$ images of NGC 1427A at 3.6 $\mu$m, 8 $\mu$m, {\it HST} H$\alpha$, and rest 28.2 $\mu$m (H$_2$ 0-0 S(0) transition) wavelengths. North is up and East is left. The colors follow the visible spectrum where blue represents the faintest emission and red the brightest. The center of the galaxy is show by the black cross. The H$\alpha$ image has been smoothed to better reveal the HII regions. The red box is the spectral extraction region used to generate the spectrum for this galaxy. The 3.6$\mu$m image shows the arc-like morphology of this galaxy. Both the H$\alpha$ and 8 $\mu$m image show star forming regions arranged along this arc. The 28.2 \mic image contains both warm dust continuum and \htwo S(0) emission. The brightest regions at 28.2 \mic are associated with areas of the most intense star formation in the galaxy. \label{n1427aimg}}
\end{figure*}

NGC1427A is a dwarf irregular galaxy within the Fornax cluster with arc-like stellar morphology. The galaxy has a moderate line-of-sight velocity with respect to the cluster mean of 650 km s$^{-1}$ (reported by NED). We adopt a distance of 20.0 Mpc to the cluster measured using the surface-brightness fluctuation technique \citep{blakeslee09}. The galaxy is at a projected separation of 130 kpc from the cluster center. We present the {\it HST} H$\alpha$, {\it Spitzer} 3.6 $\mu$m, 8 $\mu$m, and rest 28.2 $\mu$m images in Figure \ref{n1427aimg}. The 28.2 \mic image is presented instead of a 17 \mic image because no significant emission was detected at 17 $\mu$m. The 28.2 \mic image wavelength is centered on the \htwo S(0) line and also contains dust continuum emission associated with star formation. We did not subtract the dust continuum due to the low signal-to-noise detection of the S(0) line in each spaxel. The {\it HST} H$\alpha$ image has been smoothed to reveal the HII regions within the galaxy better. The arc-like star formation morphology seen at 8 \mic and H$\alpha$ and filamentary star forming regions seen at the western side of the galaxy at 8 \mic cannot be explained purely by a tidal mechanism and must require ram-pressure stripping. A tidal mechanism would affect the entire galaxy, but star formation is restricted to the periphery of the galaxy where the interaction between the ISM and the ICM is the greatest. The bulk of the 28.2 \mic emission is mainly centered on the part of the star-forming arc on the south side of the galaxy in an area where there is comparatively much more star formation than the rest of the galaxy. Other sources of 28.2 \mic emission within the galaxy seem to coincide with the peaks in 8 \mic emission.

\par
\begin{figure*}
\epsscale{0.75}
\plotone{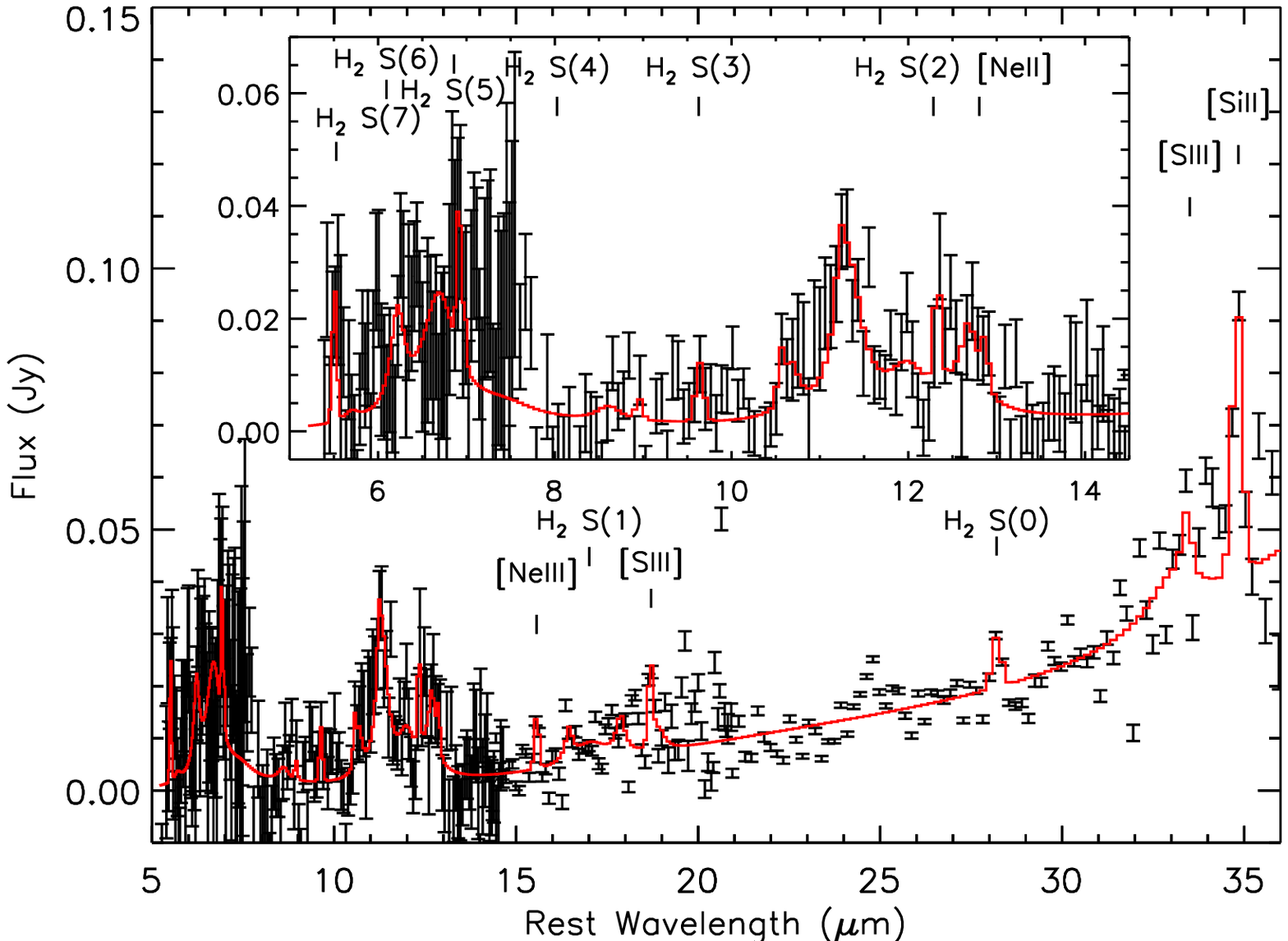}
\plotone{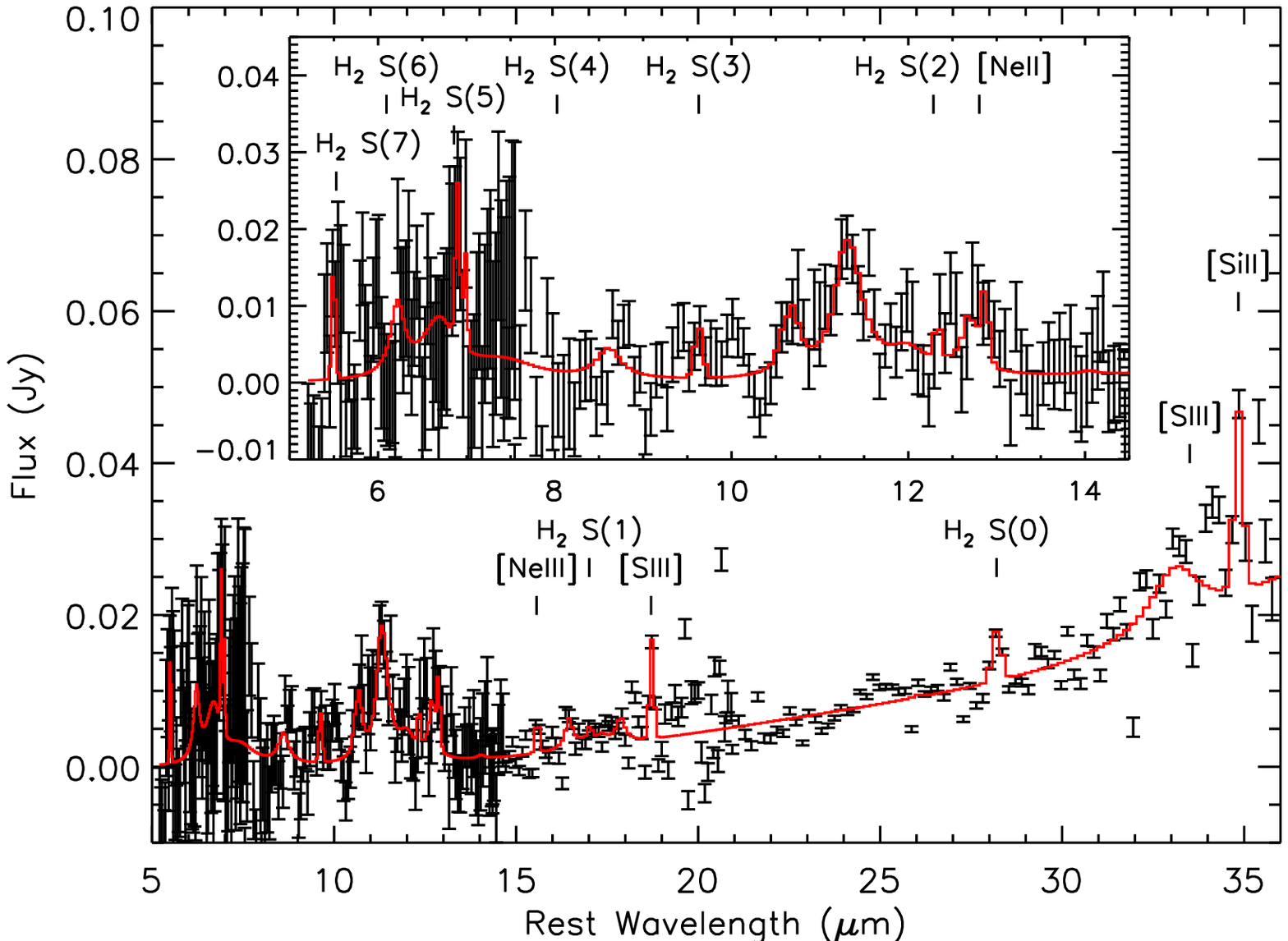}
\caption{Infrared spectra of NGC 1427A extracted from the galaxy extraction region (top) and the high IR flux region (bottom). There is only a significant detection of warm H$_2$ is the S(0) transition in both regions. The S(1) line is noticeably absent suggesting the temperature of the warm \htwo component is lower than what is generally expected and/or the ortho-to-para ratio is lower than 3. \label{n1427aspec}}
\end{figure*}

We search for signatures of warm H$_2$ within this galaxy by extracting spectra from two different regions;  the dimensions and locations of the extraction regions are shown in Figure \ref{n1427aimg}. We extract spectra from a region that covers most of the galaxy and one that focuses primarily on the high IR flux region within the galaxy.  The extracted spectra for each region are shown in Figure \ref{n1427aspec}. The S(0) line is clearly detected in both regions, but there is no hint of the S(1) line; this galaxy is the only one in our sample without a firm detection of the S(1) line. In comparison, the S(1) line is the strongest \htwo line in units of flux density observed in the other three galaxies. Because the signal-to-noise ratios of the spectra for this galaxy are low, it is worth putting this result in context. The 3-$\sigma$ upper limit to the S(1) to S(0) line ratio is unusually low with a value $\le$ 0.4. The typical values of the line flux ratio measured in other galaxies in our sample fall in the $2-6$ range, so this upper limit is a factor of 5 less than the lowest ratio we measure in our sample. 

\par
Given that our sample is relatively limited, it may not be unusual to find systems with such a low line ratio in a larger sample with measured \htwo rotational line fluxes, such as the SINGS galaxies \citep{roussel07}. The SINGS sample has a median line flux ratio of $3.1,$ which falls well within the range of values observed in the remaining three galaxies in our sample. However, there are three SINGS systems with unusually low S(1)/S(0) ratios (i.e. $< 1.0$): NGC 24, NGC 1705, and NGC 4552. All three of these systems have low ortho-to-para ratios ($0.5 < OPR < 1.5$) with warm gas component temperatures ($78\textrm{K} < T < 127$K) on the low end of the observed distribution in all sample galaxies. Two of these galaxies, NGC 24 and NGC 1705, are dwarfs and also have comparably low S(1)/S(0) ratios of $0.3-0.4.$ The fact that we do not see the S(1) line in NGC 1427A can mean two things: (1) The warm \htwo within NGC 1427A is cooler than the gas found in the other three galaxies in our sample; and/or (2) The ortho-to-para ratio (OPR) for the \htwo gas is lower than our assumed value of 3. Typically, the critical densities of the S(1) and S(0) transitions are low enough that they are thermalized in most galaxies. However, if the gas density is sufficiently low, the even and odd states of \htwo remain decoupled and the ortho-to-para ratio of the \htwo gas is essentially set to the value of the gas when it initially formed. \cite{roussel07} suggest in their work that this may explain the low observed OPR values in the three galaxies in their sample.
\par
\begin{figure}
\epsscale{1.2}
\plotone{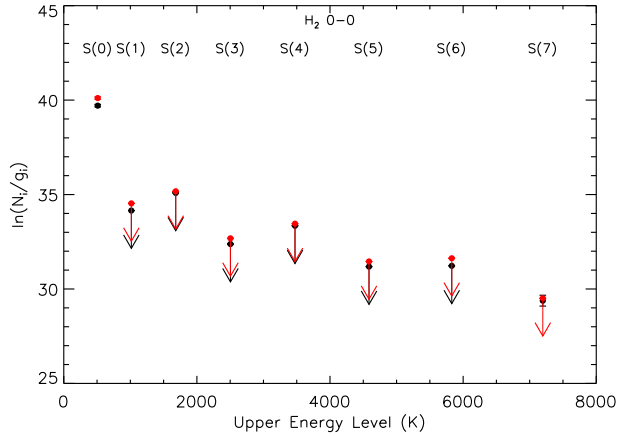}
\caption{\htwo excitation diagram for NGC 1427A. The black and red points represent the values obtained from the galactic and high IR flux extraction regions, respectively. All upper limits are 3$\sigma$ limits. In both the galactic and high IR flux extraction regions, we have a significant detection of the S(0) line. The S(7) line is also detected in the galactic extraction region, but no other \htwo lines are detected in the high IR flux region. There are insufficient line detections to reliably determine the temperature of the warm \htwo gas. The non detection of the S(1) may be the result of either lower than usual warm component temperature and/or a lower OPR value for the gas. \label{n1427aexcite}}
\end{figure}

We construct a \htwo excitation diagram, Figure \ref{n1427aexcite}, for the two regions from which the spectra have been extracted. The solid angles of the galactic and high IR flux regions are $7.96\times10^{-8}$ and $3.71\times10^{-8}$ sr, respectively. The column density values shown in the diagram assume a nominal OPR of 3. We can place constraints on the total warm \htwo mass detected within the two extraction regions by using the detection of the S(0) and the $3\sigma$ detection limit for the S(1) line. To explore the effects of the OPR, we compute \htwo mass estimates for corresponding values of 1.5 and 3 using a single temperature model. The results of these fits are given in Table \ref{gasmass}. The fit within the galactic extraction region yields a temperature upper limit of 91K and a \htwo column density lower limit of $1.2\times10^{20}$ cm$^{-2}$. Similarly, the high IR flux region is best fit by a temperature of $< 91$K and an \htwo column density of $>1.7\times10^{20}$ cm$^{-2}.$ The temperature is a few tens of Kelvin cooler than the warm component temperature observed in the other three galaxies in our sample, but is consistent with the values derived from galaxies exhibiting low S(1)/S(0) ratios in the SINGS sample. This yields a \htwo mass lower limit of $4.7\times10^7$ M$_\odot$ and $5.9\times10^7$ M$_\odot$ for the high IR flux and galactic regions, respectively. These results indicate that most of the \htwo mass is concentrated within the high IR flux region where there is significant star formation. For an OPR of 1.5, we obtain a upper limit to the temperature of 104K and a lower limit to the column density of $6.6\times10^{19}$ cm$^{-2}$ for the galactic region. This in turn translates into a lower limit on the warm \htwo gas mass of $3.4\times10^7$ M$_\odot,$ which is a 40\% reduction in mass compared to the OPR=3 case. For the high IR flux region, we obtain the corresponding temperature upper limit and column density lower limit of 103K and $1.0\times10^{20}$ cm$^{-2}.$ This yields a warm \htwo mass upper limit of $2.4\times10^{7}$ M$_\odot.$  There is a possibility that even lower OPR values may apply to the \htwo gas within NGC 1427A. This would push the upper limit of the temperature even higher than what has been found with our analysis while reducing the lower limit on the total mass. 

\par
Without a firm detection of the S(1), S(2), and S(3) lines, it is impossible to constrain the OPR ratio. We will conservatively assume that the lower limit to the \htwo mass of NGC 1427A to be $3.4\times10^7$ M$_\odot.$ To set a reasonable upper bound on the warm \htwo mass, we compute the mass using the lowest temperature of warm \htwo gas observed in the SINGS sample of 78K. It is worth noting that this temperature is observed in NGC 1705, which is one of the dwarf low OPR galaxies discussed above. This temperature choice yields a value of $1.3\times10^8$ M$_\odot.$ We compare this with the published value of HI mass of $3.3\times10^{9}$ \ms \citep{koribalski04}. The ratio of warm \htwo to HI has lies in the range $\sim 0.01$ to $\sim 0.04.$ This limit is lower than what has been observed in 97073 and NGC 4522, suggesting that there is nothing unusual about the warm \htwo emission observed in NGC 1427A and it may just be produced by star forming activities. In the discussion, additional evidence involving the H$_2$-PAH flux ratio further substantiates this conclusion.

\subsection{Dust and Star-Forming Properties}
\label{starformation}
\begin{figure}
\epsscale{1.15}
\plotone{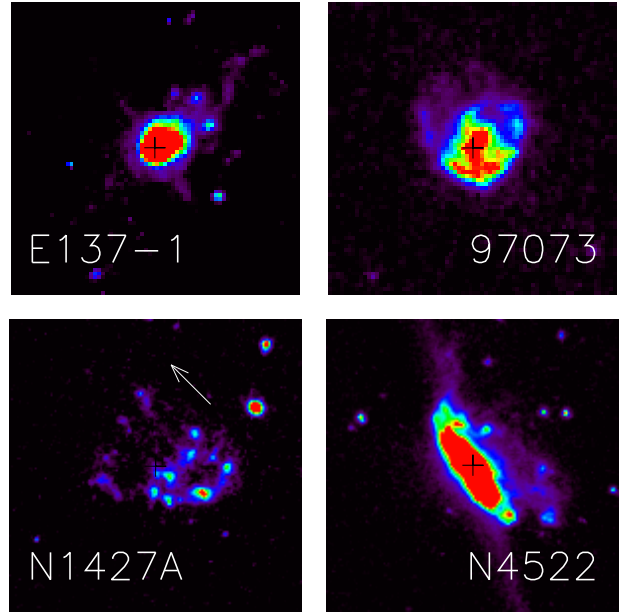}
\caption{8\mic excess images of ESO 137-001, CGCG 97-073, NGC 1427A, and NGC 4522 that reveal the aromatic dust emission from these galaxies. The images of ESO 137-001 and 97073 are $30\times30$ kpc in size, while the images of NGC 1427A and NGC 4522 are $15\times15$ kpc in size. The black crosses represent the centers of the galaxies. The image of ESO 137-001, already published in Paper 1, shows asymmetric dust emission at the galactic center and extraplanar 8\mic emission, which have been confirmed to be star-forming regions. In the case of NGC 1427A, the direction of the star formation trails is shown by a white arrow.\label{8umexcess}}
\end{figure}

\begin{figure}
\epsscale{1.15}
\plotone{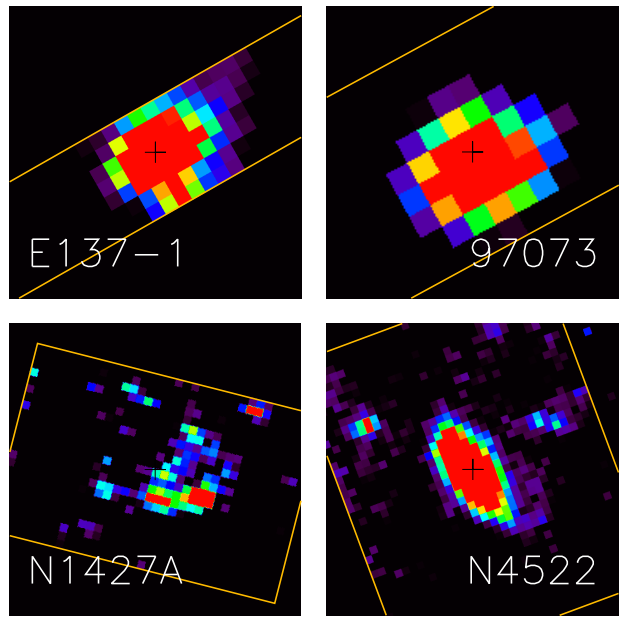}
\caption{24 \mic images of ESO 137-001, CGCG 97-073, NGC 1427A, and NGC 4522 generated from IRS spectral cubes. The images of ESO 137-001 and 97073 are $30\times30$ kpc in size, while the images of NGC 1427A and NGC 4522 are  $15\times15$ kpc in size. The orange lines show the IRS coverage and the corresponding extents of the 24 \mic images. The 24 \mic emission is produced by warm dust emission associated with star formation. The observed intensity distribution matches well with excess 8 \mic emission when the lower resolution and spatial sampling of the 24 \mic image are taken into account. \label{24umimg} }
\end{figure}

In this section, we study the star forming properties of the galaxies within our full sample and present the results of individual galaxies in the sections below. All galaxies show significant star forming activities and marked asymmetries in their distribution of star forming regions. We will not discuss the star forming properties of ESO 137-001 as it has already been discussed in detail in Paper 1. To understand star formation better in these ram-pressure stripped galaxies, we present the 8 \mic excess images in Figure \ref{8umexcess} and IRS-derived 24 \mic images in Figure \ref{24umimg}.  Even though both bands trace star formation, their utility to our analysis varies. 8 \mic emission offers a high resolution view of the morphology of star forming activities, but it is a poor indicator of the star formation rate. 24 \mic emission has poorer spatial resolution compared to 8 \mic, but it is well calibrated to provide accurate measures of star formation rates within the galaxies in question. Therefore, these bandpasses offer a complementary view of star forming activities.

\par
Thus, we have two goals. The first is to probe any effects of the ICM on star formation by comparing the morphology at 8 $\mu$m with the understanding of the stripping process from the H$_2$ behavior. The second is to see if the overall level of star formation in these galaxies is suppressed. This second issue has been examined for other cluster galaxies by \cite{tyler13}. They find that the majority of star-forming galaxies in dense clusters fall close to the field galaxy SFR/mass relation. That is, if one takes the field relation and the width of the field distribution around this relation (FWHM $\sim$ 1.2 dex, \cite{brinchmann04}), only a small minority of cluster galaxies fall significantly below expectations for field galaxies of similar mass. 

\subsubsection{CGCG 97-073}
We confirm previous observations that the majority of the star formation is confined to one side of the galaxy, specifically the side opposite the H$\alpha$ tail \citep{gavazzi01}. The arc-shaped star forming region observed by \cite{gavazzi01} at H$\alpha$ (see Figure \ref{97073img}) is also seen in our 8 $\mu$m excess image of the galaxy. In fact, the \ha and 8 \mic emission are virtually identical with the exception of the 50 kpc tail and a northern spur observed at H$\alpha$. Significant 24 \mic emission is also detected from this galaxy. The 24 \mic emission is offset towards the direction with significant star formation and the emission is centered on the area where the greatest amount of emission is seen in both H$\alpha$ and 8 \mic. Our observations strengthen the original hypothesis proposed by \cite{gavazzi01} that the galaxy is interacting with the ICM on its southern edge and that ram-pressure is compressing the molecular gas on that edge and promoting increased star formation along an arc-like star forming region. This is further supported by the lack of a similar feature at 3.6 $\mu$m. No obvious extraplanar star-forming regions or dust trails are seen at 8$\mu$m like those observed in ESO 137-001. However, it is important to note that the galaxy and its orbit may be highly inclined to our line-of-sight due to its asymmetric morphology and high line-of-sight velocity with respect to the cluster mean (v$_{gal} = 680$ km s$^{-1}$) compared to the cluster velocity dispersion of 891 km s$^{-1}$ determined by \cite{cortese04}.
\par
We measure the \emph{MIPS} 24 \mic flux for the two spectral extraction regions for the galaxy. We use the same techniques discussed in Paper 1 to calculate the equivalent \emph{MIPS} 24 \mic signal from the extracted IRS spectra. However, we use a slightly different error estimation technique to quantify the error in the flux measurement because our previous error estimation technique was far too conservative. The random errors derived from the \emph{CUBISM} generated spectra do not reflect the fluctuations present in the IR continuum due to systematic errors that arise from the reduction. We attempt to quantify properly the error by following these steps: (1) We focus on the $20-30$ \mic spectral range, as that is roughly the \emph{MIPS} 24 \mic bandpass, and we remove any spectral lines present, namely the 28.2 \mic S(0) line; (2) We carry out a polynomial fit to the spectrum and subtract the fit from the spectrum to remove the smooth component associated with the continuum; and (3) We estimate the scatter about the mean by computing the standard deviation of the subtracted spectrum. We take the error for each flux data point within the $20-30$ \mic range to be the scatter computed in step 3. The calculation of the \emph{MIPS} 24 \mic flux involves taking the weighted average of the $20-30$ \mic spectrum using the MIPS relative response curve as weights. We calculate the error in the 24 \mic flux value accordingly using the relative response curve weights as given in the following equation:
\begin{equation}
\sigma_{f24} = \left[\frac{1}{\sum_{j}c_j}\sum_i (c_i \sigma_{f,i})^2\right]^\frac{1}{2}
\end{equation}
where $\sigma_{f24}$ is the error in the 24 \mic flux, $c_i$ is the MIPS relative response value computed at each wavelength value in the IRS spectrum, and $\sigma_{f,i}$ is the flux error for each IRS data point, which we take to be the scatter value computed above. We apply this method to all of the galaxies in our sample.
\par
We measure 24 \mic fluxes of $18.8\pm0.3$ mJy and $0.30\pm0.09$ mJy for the galaxy and tail extraction regions of 97073, respectively. To determine the star formation rate (SFR) of the galaxy, we use both the \ha and 24 \mic flux measurements to account for obscured and unobscured star formation. \cite{ip02} measure an \ha flux of $1.55\times10^{13}$ ergs s$^{-1}$ cm$^{-2}.$ It is worth noting that this measurement also includes the [NII] lines because the \ha flux was measured through narrowband photometry. However, for irregular galaxies the [NII]/\ha ratio is very low ($< 0.1$) \citep{kennicutt83}, so we do not correct for this. Using the calibration derived by \cite{calzetti10}, we derive a SFR of 1.2 \msyr. Next, we determine if the galaxy's SFR is unusual in any way for its stellar mass. We estimate the stellar mass using the stellar mass-to-light ratio from the $K$-band magnitude and $B-V$ colour of the galaxy along with the \cite{bell03} mass-to-light ratio relation. 97073 has a $K$-band apparent magnitude of 12.99 and a $B-V$ colour of 0.43 \citep{gavazzi96}. This translates to a mass-to-light ratio of 0.71 and a corresponding stellar mass of $8.7\times10^{9}$\ms. We calculate a specific star formation rate (SSFR), log(SFR/M$_\star$ [yr$^{-1}$]), of $-9.9.$ This falls within the normal star forming sequence for field galaxies \citep{tyler13}. This could mean two things: (1) The cluster environment has not yet had time to quench the star formation of this galaxy; (2) Ram-pressure has induced star formation that has increased the SSFR of the galaxy as evidenced by arc-like star forming region observed in the galaxy. For a further discussion on this topic see Section \ref{starformationimpact}. 

\subsubsection{NGC 4522}
Like ESO 137-001, NGC 4522 is a spectacular example of how ram-pressure can strip dust and induce extraplanar star formation. In Figure \ref{8umexcess}, the 8\mic excess image shows evidence for significant dust stripping. The 8 \mic dust disk is much smaller than the stellar disk, hinting that dust at larger radii has already been blown off due to ram-pressure stripping. This is consistent with the outside-in picture of how ram-pressure first strips the least dense gas at larger radii as it is not as well-bound and then moves inwards as the ram-pressure strength increases. On the western side of the galaxy, there is clear evidence for dust being blown out of the disk. The most spectacular feature is the one on the southwest end of the galaxy where the largest amount of extraplanar excess 8 \mic emission is observed. The excess 8 \mic emission extends approximately 3.5 kpc from the mid-plane of the galactic disk, similar to the distribution of extraplanar warm \htwo emission. In addition to the diffuse dust emission at this location, we also see large clumps, presumably associated with star forming regions, as they are also seen at \ha. The diffuse extraplanar dust emission is also visible at 24 \mic, but we do not observe the clumps seen at 8 \mic and H$\alpha$. This may be due to a combination of reduced resolution and sensitivity at 24 $\mu$m. This observation clearly shows that dust is also significantly affected by ram-pressure and can exist some distance from the plane of the galaxy. Similar behavior was also observed in the case of ESO 137-001 where 8 \mic excess sources were seen past the tidal radius of the galaxy, though the dust seen here does not extend that far. The morphology of the dust also reveals the turbulent nature of ram-pressure stripping.
\par
\begin{figure}
\epsscale{1.15}
\plotone{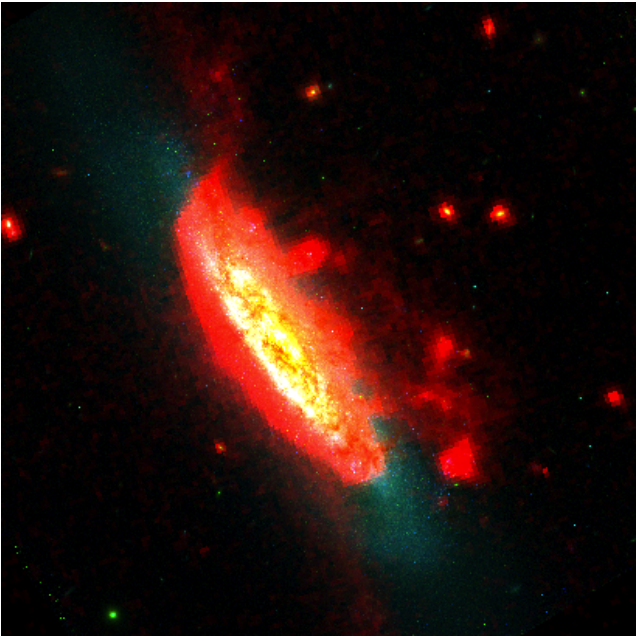}
\caption{A three-color composite image of NGC 4522. North is up and East is left. Blue is {\it HST} F435W data, green is {\it HST} F814W data, and finally red is IRAC 8 $\mu$m excess emission data. This image clearly shows that there is significant amount of dust stripping occurring in this galaxy. There are a few extraplanar 8 $\mu$m excess sources that are either offset from their optical counterparts in the direction of the ICM wind or with no optical counterparts at all in the stripped dust trail immediately west of the southern end of the galaxy. \label{n4522color}}
\end{figure}

We search for optical counterparts for the stellar-continuum subtracted 8$\mu$m emission with a particular focus on the areas with extraplanar dust. We present the color composite of {\it HST} F435W, F814W, and {\it Spitzer} 8 $\mu$m excess emission in Figure \ref{n4522color}. The large dusty clumps located immediately west of the galaxy show some blue star forming knots associated with them, but the dust emission is slightly offset to the west. There is a single 8$\mu$m source located at the western edge of the frame that has no optical counterpart. It is unclear if the source is associated with the galaxy, though it could possibly be an extraplanar star forming region. Searches within a number of catalogues (NED, SIMBAD, and VIZIER) yielded no matches to this source.
\par
We measure the 24 \mic flux within the central, NE, and SW regions and obtain $102.8\pm1$, $6.6\pm0.6,$ and $8.0\pm0.3$ mJy, respectively. Using the measured \ha flux of $4.2\times10^{-13}$ erg cm$^{-2}$ s$^{-1}$ \citep{gavazzi06} and the 24 \mic flux within the central extraction region, we obtain a star formation rate for the galaxy of 0.12 M$_\odot$ yr$^{-1}$. The extraction apertures for the \ha and 24 \mic measurements do not match exactly, but they should contain most of the flux observed at these two wavelengths. We crosscheck the measured star formation rate using 100\mic \emph{Herschel Space Telescope} photometry of this galaxy \citep{davies12}. Using the 100\mic flux-SFR relation from \cite{rieke09}, we convert a 100\mic flux density of 5.12 Jy to a SFR of 0.16 M$_\odot$ yr$^{-1},$ which is consistent with our original measurement. The 100$\mu$m-derived SFR is slightly larger because the \emph{Herschel} measurement covers the entire galaxy while we only consider the central extraction region. To be consistent with other galaxies in our sample, we only use the H$\alpha$/24$\mu$m-derived SFR for further analyses. We measure the SSFR for this galaxy to determine if the star formation rate is similar to that for a typical star forming galaxy. We calculate the stellar mass in the same manner described above. This galaxy has a $K$-band magnitude of 9.94 \citep{devereux09} and a $B-V$ colour of 0.59 \citep{schroeder96}, which yields a stellar mass-to-light ratio of 0.75 and a stellar mass of $4.4\times10^9$ M$_\odot$. This corresponds to a log(SSFR) of -10.6. This galaxy falls below the typical range of field SSFR values for its stellar mass \citep{tyler13}. This may indicate that its SFR has been suppressed through environmental processes, but it is worth noting that its SFR is not significantly lower than the observed range of typical SFR values for field galaxies. All evidence suggests that the galaxy has already experienced significant ram-pressure stripping. It is likely that this galaxy has already passed through the densest part of the ICM and is on its way out. This hypothesis is investigated further in Section \ref{starformationimpact}.

\subsubsection{NGC 1427A}
In terms of star formation and dust properties, NGC 1427A is the most unusual galaxy in our sample, as it is not a spiral galaxy but a dwarf irregular. Emission at 8 \mic and 24 \mic is clearly not as strong as in the other galaxies. A comparison of excess 8 \mic and \ha emission is particularly interesting. The {\it HST} \ha image clearly shows emission confined to an arc, meaning that the majority of star formation is occurring within the arc. This was also confirmed by \ha ground-based imaging by \cite{georgiev06}. The excess 8 \mic emission largely corroborates this view with a few significant differences. There appears to be 8 \mic emission arranged in a line directly behind the head of the arc that is not clearly seen at \ha. This may be due to a dust gradient within the galaxy that obscures part of the \ha emission. Moreover, at the eastern edge of the galaxy, finger-like excess 8 \mic emission is observed where the \ha emission is mainly dominated by point sources. These star forming regions at 8 \mic are a factor of $2-4$ fainter than the other brighter regions that are located closer to the head of the arc.  There is also a large gap spanning 2.5 kpc between the emission near the head of the galaxy and the eastern end of the galaxy, also observed at H$\alpha$. A possible interpretation may be that the star forming knots seen on the eastern end of the galaxy are associated with gas and dust that have been ram-pressure stripped, as evidenced by the wide gap between the two star forming regions and the finger-like 8 \mic excess emission. The excess 8 \mic image is also corroborated by the 24 \mic image. Due to the poorer sensitivity of the 24 \mic image, it only shows where significant star formation is occurring. It appears most of the star formation is confined to the southern side of the galaxy within the high IR flux extraction region, though some of the star formation in the eastern edge of the galaxy is also seen at 24 $\mu$m.
\par
We quantify the dust continuum emission from this galaxy by measuring the 24 \mic emission within the galaxy extraction region and high IR flux extraction region and its star formation rate. Using the methods described earlier we determine the flux values of $12.9\pm1.1$ and $7.1\pm0.6$ mJy for the galaxy and high IR flux extraction regions, respectively. Because the high IR flux region is also located within galaxy extraction region, we conclude majority of the 24 \mic flux ($\sim 60$\%) is confined to the high IR flux region. It is clear that the 24 \mic luminosity of this galaxy is considerably fainter than the other three galaxies, meaning this galaxy harbors less warm dust then the other galaxies in the sample. We also calculate the stellar mass of this galaxy in order to place these results in context. The $R$ and $I$-band magnitudes of 13.33 and 13.34, respectively, were measured by the HIPASS optical counterparts survey \citep{doyle05}. We use the \cite{bell03} $r'-i'$ mass-to-light ratio relation to estimate a stellar mass of $6.2\times10^8$ M$_\odot$, making it the lowest stellar mass object in our sample. An H$\alpha$-derived star formation rate of $0.057$ \msyr is given by \cite{georgiev06}. We convert this to a \ha luminosity to derive our own \ha/24\mic-based star formation rate using the \cite{calzetti10} relation. Assuming only foreground galactic extinction of the \ha line and the \ha-to-SFR relation used by \cite{georgiev06}, we obtain an \ha flux of $2.15\times10^{-13}$ erg s$^{-1}$ for the galaxy. This translates to a SFR of $0.065$ M$_\odot$ yr$^{-1}$, which is consistent with the value derived by \cite{georgiev06}. This means that the log(SSFR) for this galaxy is -10.0. This value falls within the distribution for field galaxies and other star-forming galaxies in dense clusters \citep{tyler13}.

\subsection{Fine Structure Lines}
\begin{figure*}
\epsscale{0.75}
\plotone{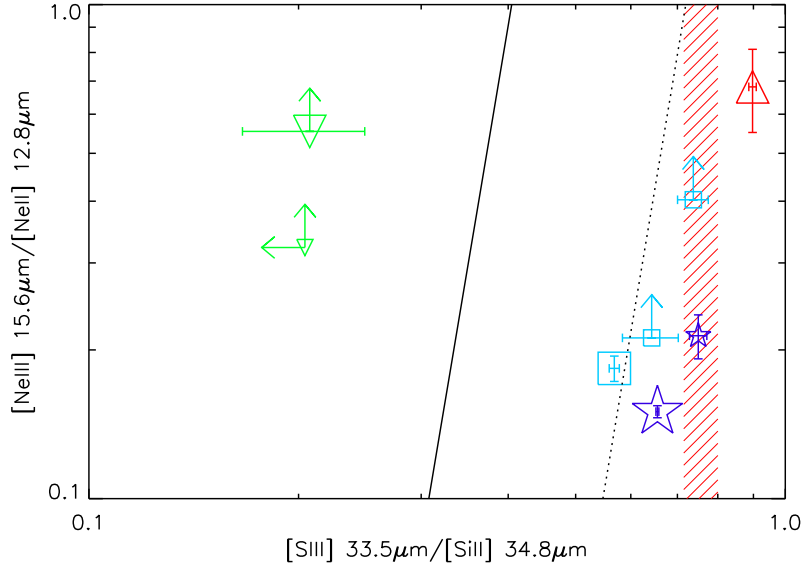}
\caption{\footnotesize Neon, sulphur, silicon fine structure line diagnostic diagram. The line ratios of these lines indicate the ionization state of the gas. The solid and dashed lines separate regions with ionization states of gas in different classes of galaxies (see \cite{dale06} for more information). Gas in Seyfert galaxies and LINERS generally falls in the region to the left of the dashed line, while gas in star forming galaxies and regions generally falls to the right of that line. There are no star forming regions found to the left of the solid line. We show the line ratio for the different regions of our galaxy sample. The red triangular point represents the values obtained from our analysis of 97073. The large symbol represents the full galaxy extraction region while the red cross-hatched box represents the allowable values for the tail extraction region because neither neon line was detected. The blue square points represent the two extraction regions in NGC 4522. The large symbol represents the nuclear region of NGC 4522 while the smaller symbols represent the two regions where ram-pressure is thought be stripping gas. The green upside down triangles represent the extraction regions in NGC1427A. The larger symbol represents the large extraction region while the smaller one represents the smaller extraction region with most of the flux. We also include data for ESO 137-001 and its tail from Paper 1 in this figure with purple star symbols. The large star symbol designates the nuclear extraction region, while the smaller ones are from the \htwo tail with IRS SL/LL coverage. With the exception of NGC 1427A, all other regions show an ionization state that is consistent with star forming regions. The ionization state of the NGC 1427A gas suggests that its gas experiences a harder radiation field or is of a higher density due to ram-pressure compression.\label{finefluxplot}}
\end{figure*}

The fine-structure line flux ratios can be used to determine the excitation mechanism for the gas, and can provide clues about the hardness of the incident radiation field. Different excitation mechanisms produce different ionization ratios and therefore fine-structure lines emitted from photodissociation regions in galaxies can be used to delineate harder excitation sources (e.g. AGN) from softer ones (e.g. star formation). \cite{dale06} in their {\it Spitzer} IRS study of Seyfert, LINERS, HII nuclei, and extranuclear HII regions found that they could isolate each type of source in a [NeIII] 15.56 $\mu$m/[NeII] 12.81 $\mu$m and [SIII] 33.48 $\mu$m/[SiII] 34.82 $\mu$m space. We carry out a similar analysis to see if there are any unusual signatures in our derived fine structure line fluxes, which may either be associated with a shock or excitation by the hard cluster X-ray radiation field. For example, if the main excitation mechanism of the gas is X-ray irradiation from the intracluster medium, one might expect more AGN-like line ratios. The fine structure line ratios for the spectral extraction regions for all of our galaxies are shown in Figure \ref{finefluxplot}. 
\par
We find that the ionization state of the gas for three of our sample galaxies, 97073, NGC 4522, and ESO 137-001, falls within the star-forming region locus of the [NeIII]/[NeII] versus [SIII]/[SiII] plot. However, NGC 1427A shows rather unusual ionization state that is similar to Seyferts and LINERs, which may be due to its close proximity to the center of Fornax clusterÕs X-ray emission. In galaxies where we see excess warm H$_2$ emission (i.e. 97073, NGC 4522, and ESO 137-001), we do not see any unusual fine-structure line ratios; there is no indication that these lines are excited by the hard X-ray emission within the cluster even in the tail regions of these galaxies. In these cases, the observed lines are most likely emitted within PDRs associated with star forming regions and not associated with the unusually excited molecular hydrogen.
\par
The unusual [SIII] 33.5$\mu$m/[SiII] 34.8$\mu$m line ratio observed in NGC 1427A could be entirely due to the faintness of the [SIII] 33.5 $\mu$m line. One way to determine if the line flux is peculiar is to compare it to a higher excitation line of the same species, the 18.7 $\mu$m [SIII] line. The typical [SIII] 18.7 $\mu$m-to-[SIII] 33.5$\mu$m ratio for the SINGS sample is 0.82 with a standard deviation of 0.27 \citep{dale06}. In the case of NGC1427A, the ratio is much higher ($> 2$). It is also worth noting that the [SIII] 18.7 $\mu$m emission is not centered on, but is slightly offset inward from, the brightest star forming region. This suggests that the faintness of the [SIII] 33.5 $\mu$m is likely real and is not likely a signal-to-noise issue. This could be potentially explained by two different effects: a hard UV radiation field and/or increased electron densities. While the former seems likely due to galaxy's proximity to the center of Fornax where the density of the hot ICM is the greatest, one can compare the [NeIII]/[NeII] ratio, which is a good indicator of hardness of radiation, with galaxies with known hard UV fields such as blue compact dwarfs (BCDs). For a select sample of low metallicity BCDs, \cite{wu08} find typical ratios of $\sim3-5,$ which is consistent with the lower limit set by our measurements. The lower [SIII]/[SiII] ratio can also be explained by increased electron densities. A ratio of $>2$ is consistent with densities $>10^3$ cm$^{-3}$ \citep{rubin89}. This galaxy already exhibits significant compression through ram-pressure, which could lead to the enhanced densities. Based on the available information, it is not possible to distinguish between these two scenarios.
\par
The blended feature of [OIV] 25.9$\mu$m and [FeII] 26.0$\mu$m has been detected in most spectra with the exception of those from NGC1427A. 
The weakness of the higher ionization [NeIII] 15.6$\mu$m line makes [FeII] the most likely line detected in the blended feature. As discussed in \cite{cluver10}, the [SiII]/[NeII] and the [FeII]/[NeII] line ratios can be used as a diagnostic for shocks. For our sample, typical values of [SiII]/[NeII] fall within a narrow range of 0.87 to 1.35 whereas the [FeII]/[NeII] ratio ranges from 0.05 to 0.20, which are similar to values observed by \cite{cluver10}. Typical SINGS galaxies that are not experiencing ram-pressure stripping have [FeII]/[NeII] ratios lower than 0.05 \citep{wong14}. Although the elevated values suggest the presence of shocks, they cannot be explained entirely by either dissociative J-shock \citep{hollenbach89} or fast shock \citep{allen08} models. J-shock models for density ranges of $10^3-10^4$ cm$^{-3}$ for a broad range of velocities produce [SiII]/[FeII] $\sim 1.$ Fast shock models on the other hand produce large [SiII]/[NeII] $\sim 10$ and [FeII]/[NeII] $\sim 1$ ratios \citep{cluver10}. A solution to this mismatch is the depletion of either Si and Fe through shocks, which has been posited by both \cite{cluver10} and \cite{wong14}. 

\section{Discussion}
\label{discussion}
The goal of our study was to test the hypothesis that warm \htwo emission can be produced through an interaction between the ICM and an infalling galaxy's ISM, and that this emission can be used as an effective tracer of ram-pressure stripped gas. Consequently, galaxies showing strong signs of on-going ram-pressure stripping were chosen for this survey. Warm \htwo emission was detected in all four galaxies at varying degrees of strength and in some cases it was seen well outside the galaxy.  In this section, we ascertain the impact on the star forming properties of galaxies due to ram-pressure, determine if the \htwo emission is anomalous and if so determine its excitation mechanisms, and how effective it may be in tracing on-going ram-pressure stripping.

\subsection{The Impact on Star Formation due to Ram-pressure Stripping}
\label{starformationimpact}
\begin{figure}
\epsscale{1.2}
\plotone{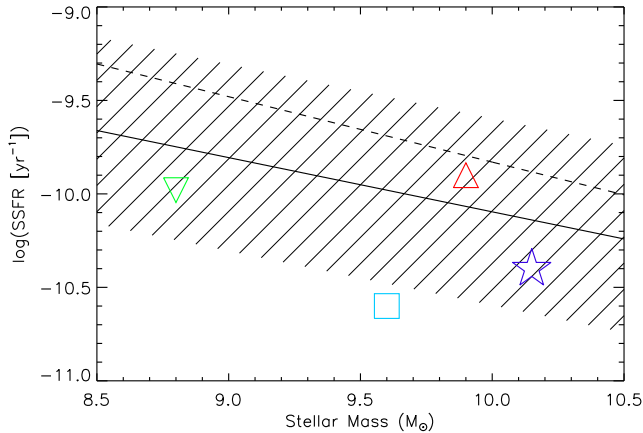}
\caption{Specific star formation rate as a function of galaxy stellar mass. The red triangle, blue square, green upside down triangle, and purple star represent 97073, NGC 4522, NGC 1427A, and ESO 137-001, respectively. The solid line is the mode of the specific star formation rate obtained from \cite{tyler13} for field galaxies, while the hashed region represents the 1$\sigma$ spread in the values. Three out of four of our galaxies fall within the typical values observed by \cite{tyler13}. NGC 4522 shows signs of suppression. The dashed line is a similar relation obtained from \cite{salim07} where the star formation rate was measured using UV data using galaxies dominated purely by star formation with no AGN contribution. The offset between the \cite{salim07} and \cite{tyler13} could be explained from systematics arising from the different measurement techniques.
\label{ssfrplot}}
\end{figure}

To determine the effects of ram-pressure on the star formation, we place our measurements of the star forming properties of the galaxies in a broader context by comparing with other studies of cluster and field galaxies. It is clear that ram-pressure is having a significant effect on the star forming properties of the galaxies simply from the observed morphological differences compared to undisturbed galaxies, such as displaced dust disks, H$\alpha$ trails, and arc-like star-forming regions. To test whether the overall star-forming rates are also affected by ram pressure, we compare to measurements of other field galaxy samples. We present the specific star formation rates as a function of stellar mass in Figure \ref{ssfrplot} for all four galaxies in our sample using H$\alpha$ and 24\micron\ measurements to estimate the SFR and stellar mass, which were presented in Section \ref{starformation}. 

\par
In order to determine if ram-pressure stripping has significantly impacted these galaxies, we compare their SSFRs with typical values for field galaxies. While many measurements of the field galaxy star formation rates exist, we compare with \cite{tyler13} measurements due to the similarity of their methods with ours. \cite{tyler13} also employ H$\alpha$ and 24\micron\ flux measurements of galaxies to determine SFRs, and use these to estimate the typical star formation rates of local field galaxies as a function of their stellar mass. We present our results of the comparison in Figure \ref{ssfrplot} where it is seen that three out of four of our galaxies, 97073, NGC 1427A, and ESO 137-001, lie within the 1$\sigma$ spread of star formation values of typical field galaxies. The four galaxies in our sample are, as an ensemble, consistent with being drawn from the field distribution (i.e., one galaxy, NGC 4522, just beyond 1-$\sigma$ is expected and seen). For reference, we also compare with other star formation studies of field galaxies that have been done with UV and/or H$\alpha$ measurements \citep[e.g.][]{brinchmann04,salim07}. UV and H$\alpha$ measurements, which often rely on an accurate determination of extinction and the intrinsic UV colors of the underlying stellar population, are not directly sensitive to heavily obscured star formation which can significantly impact estimates of SFR. This can yield significant systematic offsets in SFR measurements obtained using different methods as seen in Figure \ref{ssfrplot} where the \cite{salim07} relation is also plotted (see \cite{boquien12} and \cite{elbaz07} for more details). However, using both H$\alpha$ and 24\micron\ measurements directly accounts for both unobscured and obscured star formation as we have done in our case. In conclusion, 97073, NGC 1427A, and ESO 137-001 show signs of healthy star formation, while NGC 4522 shows, at most, a small amount of suppression.
\par
This means that ram-pressure has not significantly altered the star forming rates of these galaxies. However, there is an added possibility that ram-pressure induced star formation may compensate for any overall global reduction in star formation, as some galaxies show signs of ram-pressure enhanced star formation on the windward side of the galaxy. There is evidence observed in Virgo galaxies, including NGC 4522, that there is an enhancement in the molecular gas fraction and a possible increase of the star formation efficiency in the windward side of the galaxies \citep{vollmer12}, which together might enhance the star formation rate. However in this same work, there are signs that stripped gas, having lost the gravitational confinement of the disk, can have lower star formation efficiency. \cite{vollmer12} explain that without the gravitational potential of the disk, shocks, as observed in this work, can increase the thermal and turbulent pressure of the gas and significantly decrease its star formation efficiency as it disperses when blown out of the galaxy. This might explain why even though we observe warm \htwo emission where molecular gas has been stripped, we do not see a commensurate increase in star formation activity in those regions. 

\subsection{Comparison with SINGS Galaxies}
\label{singscomparison}
H$_2$ is typically excited within photodissociation regions through UV fluorescence in star forming regions. This explains most of the warm H$_2$ emission detected in the nearby SINGS galaxy sample \citep{roussel07}, which we use as a reference. However, our goal is to detect anomalous H$_2$ excitation that is likely associated with ram-pressure stripping. Unfortunately, it is impossible to determine the source of \htwo excitation from just the lowest ground state rotational transition line ratios because they have low enough critical densities that they are typically thermalized \citep{roussel07}. To disentangle the excitation mechanism for the H$_2$ emission, we use two different metrics that use additional information to determine if the H$_2$ emission is anomalously higher than what is produced by just star formation. 
\par
\begin{figure*}
\epsscale{0.75}
\plotone{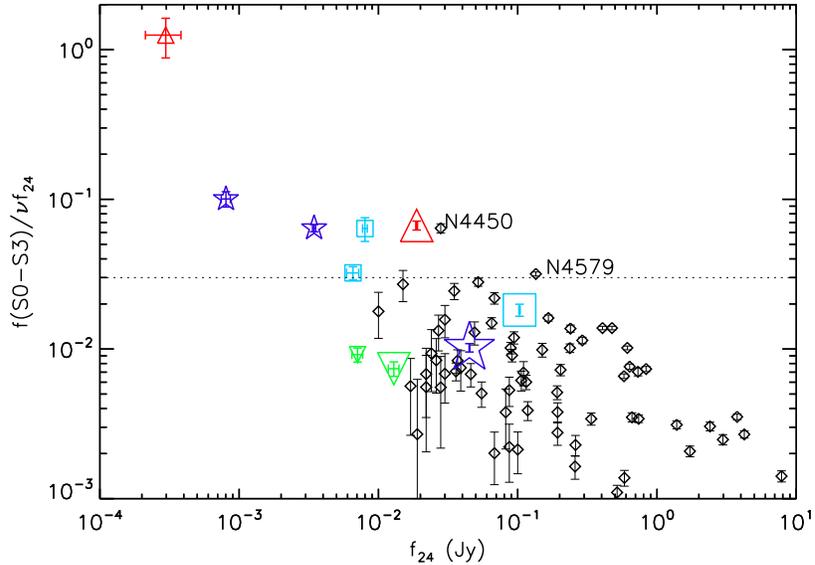}
\caption{ \footnotesize Comparison of ground state rotational H$_2$ line flux (S(0) thru S(3) transitions) to 24 $\mu$m flux as a function of 24 $\mu$m flux for the galaxies investigated in our IRS sample. In some cases where there were no detections of the S(1), S(2) and/or S(3) lines, only the detected lines are included in the \htwo flux. The open diamond points are galaxies from the SINGS sample with warm H$_2$ detections \citep{roussel07}. The significant outliers in \cite{roussel07}'s sample, NGC 4450 and 4579, are marked in the figure. The warm H$_2$ in these galaxies is thought to be heated by shocks or X-ray irradiation, as the H$_2$ line fluxes cannot be adequately explained by star formation or supernova remnants. The red triangular point represents the values obtained from our analysis of 97073. The large symbol represents the full galaxy extraction region while the smaller symbol represents the tail extraction region. The blue square points represent the two extraction regions in NGC 4522. The large symbol represents the nuclear region of NGC 4522 while the smaller symbols represent the two regions where ram-pressure is thought be stripping gas. The green upside down triangles represent the extraction regions in NGC1427A. The larger symbol represents the large extraction region while the smaller one represents the smaller extraction region with most of the flux. We also include data for ESO 137-001 and its tail from Paper 1 in this figure with purple star symbols. The large star symbol designates the nuclear extraction region, while the smaller ones are from the \htwo tail. The one with the lower flux ratio is the Tail (SL/LL) region, while the larger value is from the Far Tail (LL-only) region. In Paper 1, we showed that the warm \htwo tail in ESO 137-001 was not excited solely by star formation. This figure confirms that the warm H$_2$ observed in 97073 and the ram-pressure stripped regions of NGC 4522 are not heated solely by star formation and must have some other excitation mechanism. The tail region of 97073 is particularly striking as it has a ratio that is at least an order of magnitude higher than the regions in any other galaxy.\label{singscomp}}
\end{figure*}

The first metric, presented in Paper 1, is the ratio of \emph{MIPS} 24 $\mu$m flux, a proxy for star formation, and the sum of the line fluxes of the ground-state rotational H$_2$ lines S(0) thru S(3), a proxy for total H$_2$ line emission. We apply our metric to all of the SINGS galaxies in the \cite{roussel07} sample and show the results in Figure \ref{singscomp}. The error bars shown only include the error in the \htwo line fluxes and not the 24 \mic fluxes because 24 \mic fluxes typically have very small errors. Almost all of the galaxies have a ratio of less than 0.03, which we define as the threshold for anomalous \htwo emission. \cite{roussel07} highlight NGC 4450 and NGC 4579, two galaxies that are significant outliers in this relation, because their H$_2$ excitation cannot be explained by star formation or supernova remnants alone. They evoke either X-ray irradiation or shocks through cloud collisions as a possible explanation for the anomalous values. 

\par
We calculate the values for the metric for all of our galaxies and plot the values in Figure \ref{singscomp}. H$_2$ fluxes are obtained from line fits of the extraction regions explored in this paper and Paper 1. If the detections of the S(1), S(2), and/or S(3) lines are not significant, we do not include them in the \htwo line flux sum. The error bars show both the errors in the 24 \mic fluxes and the root-sum-squared flux for the ratio that includes both the error in the \htwo line flux and the 24 \mic flux. The values obtained from regions that represent the galaxies as a whole are plotted with larger symbols, while the other regions of interest are plotted with smaller symbols. The individual symbols for each galaxy are explained in the figure captions. It is clear that for 97073, ESO137-001, and NGC4522 the extraction regions where stripping is occurring or one sees a tail have metric values that are outliers and therefore have significantly enhanced warm \htwo emission that cannot be explained simply by star formation. The tail region of 97073, represented by the small red triangular symbol, is definitely an extreme example as it has significantly more warm \htwo excess emission than any other galaxy in the sample. In fact, the offset \htwo emission seen in 97073 strongly suggests that \htwo is being dissociated from the head of the galaxy where significant star formation is occurring and reforming some distance behind the area of intense star formation in an excited state. Furthermore, a significant fraction of its \htwo mass is in the warm state.  In terms of anomalous metric values, the far tail region of ESO 137-001 comes in as a distant second, followed by the SW region of NGC 4522. If one looks at metric values for the galactic extraction regions themselves, ESO 137-001 and NGC 4522 fall within the normal values representative of star forming galaxies in the SINGS sample. 97073, on the other hand, has enhanced emission even within the galaxy. NGC 1427A does not have any enhanced \htwo emission within either the galaxy or the high IR flux regions. In fact, it has the lowest values of the metric compared to all other galaxies and exhibits unusual \htwo emission characteristics where only the lowest energy rotational line S(0) is detected.
\par
\begin{figure*}
\epsscale{0.75}
\plotone{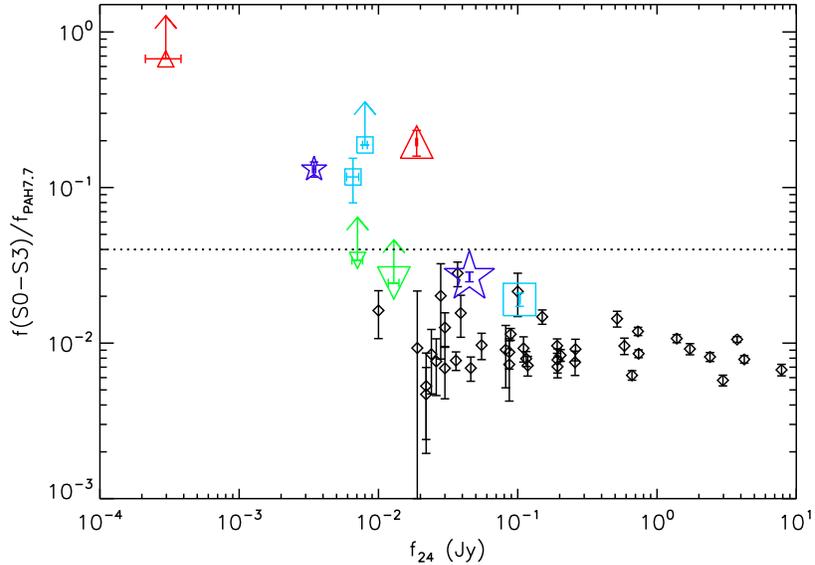}
\caption{Comparison of ground state rotational H$_2$ line flux (S(0) thru S(3) transitions) to 7.7 $\mu$m PAH feature flux as a function of 24 $\mu$m flux for the galaxies investigated in our IRS sample. In some cases where there were no detections of the S(1), S(2) and/or S(3) lines, only the detected lines are included in the \htwo flux. The symbols are the same as those used in Figure \ref{singscomp}. The Far Tail region in ESO 137-001 was not included because the IRS spectrum does not cover shorter wavelengths that included the 7.7 $\mu$m PAH feature. The open diamond points are values obtained for star forming regions from SINGS galaxies from \cite{roussel07}. The SINGS points have an average H$_2$ to PAH ratio of $\sim0.01.$ The dotted line represents the maximum ratio that can be explained by PDRs within normal star-forming galaxies and dwarfs \citep{ogle10,guillard12}. Larger ratios are most likely explained by shock excitation of H$_2.$
\label{singspahcomp}}
\end{figure*}

The second metric more directly tracks star formation powered photodissociation regions (PDR) as being the location of the warm H$_2$ emission. In recent work of galaxies with anomalous \htwo emission, the \htwo line flux versus 7.7$\mu$m PAH flux ratio has been an excellent discriminator of anomalous \htwo emission \citep{appleton06,roussel07,ogle10,guillard12b}. \cite{ogle10} empirically show through a comparison of SINGS galaxies that their molecular hydrogen emission galaxies (MOHEGs) are powered either by shocks or cosmic rays because their H$_2$-PAH flux ratios fall in the $0.04-4$ range. \cite{guillard12b}, in Figure 7 of their work, show the large range of galaxies with anomalous \htwo emission and confirm that star formation powered \htwo emission cannot explain flux ratios greater than 0.04. They further support this claim by running PDR models using existing code \citep{lepetit06}, which were found to be consistent with other models of \htwo emission in PDRs \citep{kaufman06}. We present the flux comparison for our measurements in Figure \ref{singspahcomp}. It is worth noting that the far tail region of ESO 137-001 where this ratio would be the highest in that system is not included due to the lack of short wavelength IRS coverage. We also plot the ratios for star forming SINGS galaxies using the 7.9$\mu$m and H$_2$ line flux measurements carried out by \cite{roussel07}. It is clear that the star forming SINGS galaxies form a narrow band in  H$_2$/PAH flux ratio with typical values slightly less than 0.01. The results of our measurement are consistent with our conclusions from the use of our first metric. The regions that are experiencing stripping all fall above the 0.04 flux ratio cut with the exception of NGC1427A, which is thought not to be experiencing significant stripping. 97073 remains the most dramatic example falling well above the relation with almost 60 times  the flux ratio of star forming SINGS galaxies. NGC 4522 shows modestly elevated flux ratios with almost 10 times the SINGS average. This is slightly higher than similar measurements made by \cite{wong14} of the same galaxy where they find an increase of $3-5$ times of the SINGS average. The difference may be attributed to different extraction regions and our inclusion of the S(3) to the \htwo flux measurement. Nevertheless, two conclusions can be drawn from this comparison: (1) star formation itself is insufficient for exciting the \htwo and (2) We observe a sequence of galaxies with varying levels of \htwo excitation, which may be due to the current evolutionary state of the galaxy; this is discussed further below.

\subsection{The Source of Warm H$_2$ Excitation}
As  discussed in the introduction, a number of potential excitation mechanisms for \htwo beyond star formation were reviewed. To get a better understanding of potential excitation mechanisms, we can estimate the timescale of warm \htwo emission if there was no {\it in-situ} energy source. Using the ideal gas law for a diatomic molecule, the time scale $\tau$ can be approximated to be
\begin{eqnarray}
\tau & \sim & \frac{5}{2}N_{H_2} k T_{H_2}/L_{H_2}
\end{eqnarray}
where $N, T,$ and $L$ are the number, temperature, and luminosity of \htwo molecules, respectively. For our estimation, we only consider the warm component of the \htwo since it has been characterized in the tail regions of all three galaxies with anomalously high \htwo emission. We take $L_{H_2}$ to be the sum of the luminosities of the S(0) and S(1) lines. We obtain an $L_{H_2}$ of $3.2\times10^6\:L_\odot,$ $5.6\times10^6\:L_\odot,$ and $4.6\times10^5\:L_\odot$ for the Full Tail region of ESO137-001, the Tail region of CGCG97-073, and both the NE and SW tail regions of NGC4522, respectively. The timescales of these regions correspond to 4, 8, and 8 kyrs. One can also estimate the true time scale of the \htwo emitting regions if the geometry of a galaxy's orbit and its velocities are known. ESO 137-001 is the best candidate for this measurement as its orbit is along the plane of the sky. Taking a conservative galaxy velocity of 1600 km s$^{-1}$ and a tail length of 20 kpc \citep{sivanandam10}, we obtain a time scale of its \htwo tail of 12 Myrs. A similar estimation of NGC4522 can be made, but its orbit is not along the plane of the sky. With its projected tail length of 4kpc and a galaxy velocity estimate of 1000 km s$^{-1}$ along the plane of the sky, we obtain a timescale of 4 Myrs. It is clear from our sample that the true timescales of these warm \htwo tails are much longer than the radiative timescales of the warm gas. This means that there must an {\it in-situ} source of energy that continuously heats the gas.
\par
Of the several potential excitation sources such as MHD waves, conduction, cosmic ray heating, and X-ray dissociation regions (XDRs), shock excitation seems to be the most likely explanation. We can make a morphological argument to support this view. In most cases, it is clear that the central regions of these galaxies do not show anomalous H$_2$-PAH flux ratio whereas their stripped \htwo trails show significant enhancement of \htwo emission. This is especially seen in 97073 in Figure \ref{97073mips} where there is a clear offset between the regions with significant star formation where \htwo does exist and those where warm \htwo is observed. The cluster centric sources such as MHD waves and cosmic ray heating should not discriminate where in the galaxy they deposit their energy. Moreover, these processes generally yield hotter (T$\sim 300$K; \cite{johnstone07}) \htwo temperatures. For similar reasons, conduction and XDRs cannot adequately explain the morphology. Of course, if the observed excitation is indeed due to turbulent shocks, which are induced by ram-pressure stripping, there needs to be sufficient kinetic energy dissipated to account for the observed \htwo luminosity.
\par
To calculate the amount of energy extracted from the ram-pressure stripping process, we present a toy model to estimate the loss in kinetic energy of the galaxy from ram-pressure drag as it falls into a galaxy cluster. This energy would then be injected into the stripped ISM of the galaxy and the ICM either through turbulent or viscous processes as the galaxy travels through the cluster (see \cite{roediger08} for a discussion). Ram-pressure, $P_{ram},$ introduces a drag force that decelerates the galaxy:
\begin{eqnarray}
a_{drag} & = & -\frac{P_{ram}\: A_{ISM}}{M_{gal}} \\
	& = & -\frac{\rho_{ICM}\:v_{gal}^2\:A_{ISM}}{M_{gal}}
\end{eqnarray}
where $a_{drag}$ is the deceleration experienced by the galaxy, $A_{ISM}$ is the galaxy ISM's cross-sectional area, $M_{gal}$ is the mass of the galaxy, $v_{gal}$ is its orbital velocity, and finally, $\rho_{ICM}$ is the density of the ICM at the galaxy position. If we assume the ram-pressure is not changing in relative short time scales and that the galaxy is travelling face-on to the ICM wind, we can make the following simplification to determine loss of kinetic energy:
\begin{eqnarray}
\frac{dE}{dt} & = & \frac{d}{dt}( M_{gal}\: a_{drag}\: s) \\
		   & = & M_{gal}\: a_{drag} \frac{ds}{dt} \\
		   & = & -P_{ram}\: A_{ISM}\: v_{gal} \\
		   & = & -\rho_{ICM}\: v_{gal}^3\: A_{ISM}
\end{eqnarray}
where $s$ is the distance traveled by the galaxy. The above equation can be written in more convenient units below:
\begin{eqnarray}
\frac{dE}{dt} & = & -7.8\times10^8\left(\frac{\rho_{ICM}}{10^{-27}\:g\: cm^{-3}}\right)\times \nonumber \\
	& & \left(\frac{v_{gal}}{1000\:km\:s^{-1}}\right)^3\left(\frac{r_{gal,ISM}}{10\:kpc}\right)^2\:L_\odot
\end{eqnarray}
A few observations can be made from this derivation. First, the energy loss is the strongest as an infalling galaxy approaches the cluster core where the ICM density is the highest and the infalling galaxy is travelling the fastest. Second, as the ISM of the galaxy is stripped away, the cross-sectional area decreases thereby reducing the kinetic energy loss. We calculate the expected kinetic energy loss rate for the three galaxies with anomalous \htwo emission. The calculated energy loss rate for ESO 137-001 with values obtained from Paper 1, $\rho_{ICM} = 2.2\times10^{-27}g\: cm^{-3},$ $v_{gal}=1602\:km\:s^{-1},$ and $r_{gal,ISM} = 3\:kpc,$ is $6.4\times10^8\:L_\odot.$ This is a few orders of magnitude more than the observed \htwo luminosity. Moreover, this galaxy has an observed X-ray \citep{sun06, sun10} and H$\alpha$ \citep{sun07} tails with combined luminosity in the $10^7\:L_\odot$ range. The combined power output of these tails is consistent with the maximum power produced by the dissipation of the galaxy's kinetic energy through ram-pressure stripping. A similar calculation for NGC 4522 for a parameter choice of $\rho_{ICM} = 10^{-28}\:g\:cm^{-3}$, $v_{gal} = 1500\:km\:s^{-1}$ \citep{kenney04}, and $r_{gal,ISM} = 2.4\:kpc$ (estimated from the 8$\mu$m image) yields a energy loss of $1.5\times10^7\:L_\odot.$  Finally for 97073, the calculated energy loss rate is $1.2\times10^9\:L_\odot.$ For this galaxy, $\rho_{ICM}$ of $1.1\times10^{-27}\:g\:cm^{-3}$ was obtained from \cite{mohr99} and $v_{gal}$ was estimated to be $\sqrt{3}\sigma$ where $\sigma,$ the cluster velocity dispersion, was $891$ km s$^{-1}$ \citep{cortese04}. $r_{gal,ISM}$ was estimated to be 6.1 kpc from the 8$\mu$m image. All of the observed \htwo luminosities can be explained by the energy loss by the ram-pressure drag experienced by the galaxy. What is also particularly interesting is that the calculated energy loss rate seems to track the magnitude of anomalous warm \htwo emission discussed in the previous section. This may be additional evidence for the dissipation of kinetic energy being the source of energy for the observed emission.

\subsection{Tracing Ram-Pressure Stripping with Warm \htwo Emission}
\label{trace}
We place the results of our survey in a general context by attempting to answer the question of how excess warm \htwo emission relates to ram-pressure stripping. As discussed in Section \ref{singscomparison}, three of the four galaxies show enhanced \htwo emission and all three of them show evidence for a warm \htwo tail, while two of them, ESO 137-001 and NGC 4522, show strong evidence for extraplanar star formation. The fourth galaxy shows signs of ram-pressure induced star formation but no enhanced warm \htwo emission or tail. If one arranges all galaxies in ascending order of their total gas mass, assuming ESO 137-001 has a HI mass close to the measured upper limit of $1\times10^9$ \ms \citep{sun07}, we would have NGC 4522 with the least mass, followed by ESO 137-001, 97073, and NGC 1427A. In descending order of projected distance from their respective cluster centers we have: NGC 4522, 97073, ESO 137-001, and NGC 1427A. 
\par
This suggests a possible evolutionary sequence as an explanation for the observed differences in the properties of the four different galaxies. The first stage is represented by NGC 1427A. NGC 1427A is likely at the initial stages of being stripped by ram-pressure where the ICM does not have significant strength to strip out the molecular and HI gas, as evidenced by the extremely large HI mass measurement for this galaxy \citep{koribalski04}. This is plausible because, even though NGC 1427A is very close to the center of the Fornax cluster (only 130 kpc in projected separation), the cluster is very poor and has a fairly low ICM temperature of 1.6 keV \citep{ikebe02}. Furthermore, this galaxy has a large recessional velocity with respect to the cluster mean ($\sim$ 650 km $^{-1}$), and it is also possible that its actual radial separation from the cluster core may be much larger than what is observed. This galaxy was originally chosen to be part of the sample due to its prominent arc-like star forming regions \citep{georgiev06}. It also has an unusually large atomic gas reservoir. The galaxy is most likely experiencing some ram-pressure, which is able to induce star formation by compressing gas to form stars where the ram-pressure is significant, but the pressure is not high enough to remove molecular gas or excite \htwo through the interaction.
\par
The second stage is likely represented by 97073 where the interaction between the ICM and the galaxy's ISM becomes strong enough to dissociate the molecular gas and reform it in an excited state downstream, possibly through shocks, in addition to inducing star formation along the leading edge of the galaxy. 97073 is in a moderately rich cluster that has a relatively hot ICM with a temperature of 3.6 keV \citep{ikebe02}. The images of the galaxy at H$\alpha$, and 8 \mic clearly show an arc of star formation occurring at the location directly opposite to the \ha tail. The 24 \mic image also shows intense dust continuum emission that coincides with the star forming arc. Furthermore, this galaxy's tail region exhibits the strongest warm \htwo excess of all the regions explored, while the galaxy itself exhibits an excess that is stronger than all other galactic regions considered. It is very likely that this galaxy is just beginning to experience strong ram-pressure.  Its excess 8 \mic emission is not truncated in any way, which suggests that there is sufficient molecular gas throughout the galaxy to promote star formation. It has a healthy amount of gas, both in molecular and atomic form, and it has a healthy star formation rate. The current rate of stripping cannot be as significant as ESO 137-001 because we do not see large star forming knots trailing behind the galaxy. Unfortunately, due to the alignment of the IRS slit, the true length of the warm \htwo tail was not determined.
\par
The third stage is probably represented by ESO 137-001 where a significant fraction of atomic and molecular gas has been stripped from the galaxy and a significant fraction of molecular gas is within the observed \ha and X-ray tails as discussed in Paper 1. This galaxy is also fairly close in projected separation ($\sim 280$ kpc) to the core of Abell 3627, which is the hottest cluster in our sample with an ICM temperature of 6 keV \citep{sun06}. The galaxy has an upper limit on its HI mass of $10^9$ \ms \citep{sun07} and no published values for its cold molecular gas content. The galaxy still has a relatively healthy star formation rate. It is clear from the 8 \mic excess image that the emission is not as extended as the stellar disk and the galaxy has a trail of star forming knots stretching approximately 12 kpc from the galaxy center. This trail of star forming knots is mostly embedded within the $>20$ kpc long warm \htwo tail. The warm \htwo gas within the tail cannot be explained simply by star formation. The existence of long gaseous tails, the long trail of star forming regions, and truncated dust emission suggest that this galaxy may be further along in its ram-pressure stripping phase than the previous two galaxies. 
\par
NGC 4522 possibly represents the last stage of this process. Observational signatures show this galaxy to be similar to ESO 137-001 in many ways. It has a smaller dust disk than its stellar disk. It has extraplanar dust emission albeit at a smaller scale, and similarly a $\sim 4$ kpc long \htwo tail, which has a factor of $5-10$ less mass than the \htwo tail in ESO 137-001. This galaxy is also fairly poor in gas, with comparable HI and molecular gas content, and the HI gas is known to be displaced in the direction of stripping \citep{kenney04}. All signs suggest that this galaxy is either in a similar phase to ESO 137-001 but is less affected by ram-pressure because this galaxy is in a much cooler cluster and not as dense a region of the ICM, or in a later phase where ram-pressure is becoming less significant. The morphologies of its dust and \htwo trails also suggest that it is on its way out of the cluster core. \cite{vollmer06} suggest from their dynamical modeling of VLA HI data that this galaxy must have passed through peak ram-pressure stripping 50 Myrs ago. Spectroscopic observations corroborate this finding where a K+A spectrum is seen in the outer stellar disk of the galaxy. Stellar population modelling suggests that the star formation in the outer stellar disk was quenched $\sim100$ Myr ago \citep{crowl06}. This most likely explains the suppressed star formation observed in this galaxy. This galaxy has a few peculiarities as it is approximately 1 Mpc away from Virgo's center and the the traditional view is that ICM cannot be dense enough to ram-pressure strip the galaxy significantly in a fairly low temperature cluster \citep[$\sim 2.4$ keV;][]{white00} at that distance assuming a typical velocity for a galaxy, as discussed above. But in this case, the galaxy has a fairly large recessional velocity with respect to the cluster mean ($\sim 1000$ km s$^{-1}$). While it is clear from the multiple observational signatures that the galaxy is experiencing ram-pressure stripping, this may require that the ICM be dynamic, i.e. it is moving relative to the cluster mean, and its density is enhanced, or the galaxy is traveling at very high speeds and is unbound from the cluster \citep{vollmer06}. 
\par
Our study shows that warm \htwo heated by the interaction between the ICM and ISM is not uncommon in cluster galaxies experiencing significant ram-pressure stripping. In the most extreme cases of warm \htwo emission, i.e. ESO 137-001 and 97073, there are long $40-70$ kpc tails observed at other wavelengths. There may be a strong association between the existence of an extra planar star formation/\ha emission and a warm \htwo tail. However, a greater number of detections of \htwo tails is required to confirm this association. If one assumes such an association is true, the recent discovery of more than a dozen galaxies in Coma with extended \ha knots \citep{yagi10} and UV trails \citep{smith10} may suggest that warm \htwo tails may be present in all galaxies with significant ram-pressure stripping, and may be an indicator of direct stripping of molecular gas.

\section{Conclusions}
We present the results of a study of four galaxies, known to be currently undergoing ram-pressure stripping, to detect shock-heated warm \htwo emission associated with galaxies' interaction with the ICM. The results for one of our galaxies, ESO 137-001, have already been published in Paper 1. The main results of our study are as follows:
\par
1. We detect warm \htwo emission in all four galaxies within the sample. The ground state rotational \htwo S(0) line is observed in all four of galaxies. The S(1), S(2), and S(3) lines are detected in ESO 137-001, CGCG 97-073, and NGC 4522. Warm \htwo emission is also detected in extraplanar regions of ESO 137-001, CGCG 97-073, and NGC 4522 indicating that molecular gas is present in ram-pressure stripped gas. ESO 137-001 exhibits a warm \htwo tail $> 20$ kpc in length, while NGC 4522 has one that is $\sim 4$ kpc in length. Even though a portion of the tail was observed in CGCG 97-073, the length of this galaxy's warm \htwo tail could not be ascertained. NGC 1427A exhibited no signs of a \htwo tail, but had strong \htwo emission associated with the brightest star-forming regions located on the southern edge of the galaxy.
\par
2. We measure the thermodynamic properties of the warm \htwo gas, such as temperature, column density, and total mass, in all four galaxies both within the galaxies themselves and also in other regions of interest likely associated with ram-pressure stripping. We find similar temperature distributions in ESO 137-001, CGCG 97-073, and NGC 4522. The \htwo emission could be adequately fit by a two-temperature model, one warm (T$=115-160$K) and the other hot (T$=400-600$K). In the case of NGC 1427A, we were only able to place an upper limit on its gas temperature of T$=90-105$K due to the lack of detection of the S(1) line. This anomaly could be explained by either a lower than expected ortho-to-para ratio and/or unusually cool \htwo gas. The column densities of the warm \htwo gas within the explored regions vary somewhat but fall within the $10^{19}-10^{20}$ cm$^{-2}$ range. The hot \htwo gas typically has densities that are two orders of magnitude less than the warm H$_2,$ making the warm gas the dominant mass component. The detected total masses of warm H$_2$ amongst the explored regions within all four galaxies range from $10^{6}-10^{8}$ M$_\odot,$ with CGCG 97-073 having the largest mass.
\par
3. We observe clear signs of dust stripping in at least 2 out of 4 galaxies in our 8\mic images in the form of extraplanar dust emission, which is also visible in the 24 \mic images. ESO 137-001 reveals emission by dust in a cometary morphology that is extended in the direction of the wind. In the case of NGC 4522 where we have the best spatial resolution, the turbulent nature of dust stripping is observed, as there are large knots and ripples in the stripped regions. It is possible that NGC 1427A may also be experiencing dust stripping as evidenced by long finger-like emission at 8 $\mu$m. 
\par
4. We study the star-forming morphologies of each of the galaxies and find signatures of ram-pressure induced arc-like star formation within the galaxy and/or extraplanar star formation from blown out gas. NGC 1427A is likely the most dramatic case of ram-pressure induced star formation within the galaxy, followed by CGCG 97-073, as both galaxies exhibit strong arc-shaped star forming regions. ESO 137-001 and NGC 4522 have extraplanar star forming regions that span 12 and 3 kpc, respectively.
\par
5. We measure the star-forming rates in each galaxy within our sample to see if they are suppressed within the cluster environment. Three out of four galaxies, 97073, NGC 1427A, and ESO 137-001, fall within the typical star forming sequence for field galaxies, while NGC 4522 star formation rate falls slightly more than 1$\sigma$ low. 
\par
6. We measure common fine structure lines and find that three out of four of our galaxies do not show unusual ionization states and they are generally consistent with star forming regions. However, the gas in NGC 1427A shows a very unusual ionization state that is similar to Seyferts and LINERs, which can be explained by higher electron densities or harder radiation field. The [FeII]/[NeII] ratio is elevated in our sample, suggesting potential shock excitation, but when combined with the [SiII]/[NeII] line ratio, the observed values are neither consistent with J-shock or fast shock models. This may be due to either Fe and/or Si depletion from shocks.
\par
7. We compare the measured \htwo flux with the 24 \mic flux and 7.7$\mu$m PAH flux to determine if the \htwo emission is unusual in any way compared to a control sample consisting of SINGS galaxies with warm \htwo measurement. In three of our four galaxies, we observe excess \htwo emission, especially in regions where stripped gas is present. CGCG 97-073 is a special case where both the galaxy and the tail region are outliers, and its tail region is the most significant outlier of all the regions considered. NGC 1427A is the only galaxy without anomalously high \htwo emission. 
\par
8. There may be a possible association between extraplanar \ha and \htwo emission because three of the four galaxies have coaligned \ha and warm \htwo emission.
\par
9. We hypothesize that excess warm \htwo emission may be a common feature of galaxies that are experiencing significant ram-pressure stripping and can be used as a tracer for this process. The varying degree of excess \htwo emission we have observed in our sample is most likely the result of galaxies at different stages of ram-pressure stripping (i.e. experiencing different levels of ram-pressure). 
\par
Lastly, when the \emph{James Webb Space Telescope} is launched in the near future, the mid-infrared spectroscopic window will be opened again with the integral field spectrometer on the Mid-infrared Instrument. This will allow a much larger sample of galaxies to be investigated. 

\acknowledgements
We thank Dr. Ming Sun and Dr. Lei Bai for their helpful suggestions. S. S. was supported by the Dunlap Fellowship at the University of Toronto. We thank the referee for their great suggestions to improve the paper. We also would like to thank the {\it Spitzer}/GTO team for support at the University of Arizona. This research has made use of the NASA/IPAC Extragalactic Database (NED) which is operated by the Jet Propulsion Laboratory, California Institute of Technology, under contract with the National Aeronautics and Space Administration. This research has made use of the GOLD Mine Database.


\begin{thebibliography}{}

\bibitem[Allen et al.(2008)]{allen08} Allen, M.~G., Groves, B.~A., Dopita, M.~A., Sutherland, R.~S., \& Kewley, L.~J.\ 2008, \apjs, 178, 20

\bibitem[Appleton et al.(2006)]{appleton06} Appleton, P.~N., et al.\ 2006, \apjl, 639, L51 

\bibitem[Bai et al.(2007)]{bai07} Bai, L., Marcillac, D., Rieke, G.~H., et al.\ 2007, \apj, 664, 181 

\bibitem[Bell et al.(2003)]{bell03} Bell, E.~F., McIntosh, D.~H., Katz, N., \& Weinberg, M.~D.\ 2003, \apjs, 149, 289

\bibitem[Bigiel et al.(2011)]{bigiel11} Bigiel, F., Leroy, A.~K., Walter, F., et al.\ 2011, \apjl, 730, L13

\bibitem[Blakeslee et al.(2009)]{blakeslee09} Blakeslee, J.~P., Jord{\'a}n, A., Mei, S., et al.\ 2009, \apj, 694, 556

\bibitem[Bloemen et al.(1986)]{bloemen86} Bloemen, J.~B.~G.~M., Strong, A.~W., Mayer-Hasselwander, H.~A., et al.\ 1986, \aap, 154, 25

\bibitem[Boquien et al.(2012)]{boquien12} Boquien, M., Buat, V., Boselli, A., et al.\ 2012, \aap, 539, A145

\bibitem[Boselli et al.(1994)]{boselli94} Boselli, A., Gavazzi, G., Combes, F., Lequeux, J., \& Casoli, F.\ 1994, \aap, 285, 69 

\bibitem[Boselli \& Gavazzi(2006)]{boselli06} Boselli, A., \& Gavazzi, G.\ 2006, \pasp, 118, 517

\bibitem[Brinchmann et al.(2004)]{brinchmann04} Brinchmann, J., Charlot, S., White, S.~D.~M., et al.\ 2004, \mnras, 351, 1151

\bibitem[Calzetti et al.(2007)]{calzetti07} Calzetti, D., et al.\ 2007, \apj, 666, 870 

\bibitem[Calzetti et al.(2010)]{calzetti10} Calzetti, D., Wu, S.-Y., Hong, S., et al.\ 2010, \apj, 714, 1256

\bibitem[Chanam{\'e} et al.(2000)]{chaname00} Chanam{\'e}, J., Infante, L., \& Reisenegger, A.\ 2000, \apj, 530, 96

\bibitem[Cluver et al.(2010)]{cluver10} Cluver, M.~E., et al.\ 2010, \apj, 710, 248

\bibitem[Corbelli et al.(2012)]{corbelli12} Corbelli, E., Bianchi, S., Cortese, L., et al.\ 2012, \aap, 542, A32

\bibitem[Cortese et al.(2004)]{cortese04} Cortese, L., Gavazzi, G., Boselli, A., Iglesias-Paramo, J., \& Carrasco, L.\ 2004, \aap, 425, 429

\bibitem[Cortese et al.(2007)]{cortese07} Cortese, L., et al.\ 2007, \mnras, 376, 157

\bibitem[Cortese et al.(2010)]{cortese10} Cortese, L., Davies, J.~I., Pohlen, M., et al.\ 2010, \aap, 518, L49

\bibitem[Crowl \& Kenney(2006)]{crowl06} Crowl, H.~H., \& Kenney, J.~D.~P.\ 2006, \apjl, 649, L75

\bibitem[Dale et al.(2006)]{dale06} Dale, D.~A., Smith, J.~D.~T., Armus, L., et al.\ 2006, \apj, 646, 161 

\bibitem[Davies et al.(2012)]{davies12} Davies, J.~I., Bianchi, S., Cortese, L., et al.\ 2012, \mnras, 419, 3505 

\bibitem[Devereux et al.(2009)]{devereux09} Devereux, N., Willner, S.~P., Ashby, M.~L.~N., Willmer, C.~N.~A., \& Hriljac, P.\ 2009, \apj, 702, 955 

\bibitem[Donahue et al.(2011)]{donahue11} Donahue, M., de Messi{\`e}res, G.~E., O'Connell, R.~W., et al.\ 2011, \apj, 732, 40 

\bibitem[Doyle et al.(2005)]{doyle05} Doyle, M.~T., et al.\ 2005, \mnras, 361, 34 

\bibitem[Egami et al.(2006)]{egami06} Egami, E., Rieke, G.~H., Fadda, D., \& Hines, D.~C.\ 2006, \apjl, 652, L21

\bibitem[Elbaz et al.(2007)]{elbaz07} Elbaz, D., Daddi, E., Le Borgne, D., et al.\ 2007, \aap, 468, 33

\bibitem[Fazio et al.(2004)]{fazio04} Fazio, G.~G., et al.\ 2004, \apjs, 154, 10 

\bibitem[Ferland et al.(2008)]{ferland08} Ferland, G.~J., Fabian, A.~C., Hatch, N.~A., et al.\ 2008, \mnras, 386, L72 

\bibitem[Ferland et al.(2009)]{ferland09} Ferland, G.~J., Fabian, A.~C., Hatch, N.~A., et al.\ 2009, \mnras, 392, 1475

\bibitem[Fumagalli et al.(2009)]{fumagalli09} Fumagalli, M., Krumholz, M.~R., Prochaska, J.~X., Gavazzi, G., \& Boselli, A.\ 2009, \apj, 697, 1811

\bibitem[Gavazzi \& Boselli(1996)]{gavazzi96} Gavazzi, G., \& Boselli, A.\ 1996, Astrophysical Letters Communications, 35, 1

\bibitem[Gavazzi et al.(2001)]{gavazzi01} Gavazzi, G., Boselli, A., Mayer, L., Iglesias-Paramo, J., V{\'{\i}}lchez, J.~M., \& Carrasco, L.\ 2001, \apjl, 563, L23

\bibitem[Gavazzi et al.(2003)]{gavazzi03} Gavazzi, G., Boselli, A., Donati, A., Franzetti, P., \& Scodeggio, M.\ 2003, \aap, 400, 451 

\bibitem[Gavazzi et al.(2006)]{gavazzi06} Gavazzi, G., Boselli, A., Cortese, L., Arosio, I., Gallazzi, A., Pedotti, P., \& Carrasco, L.\ 2006, \aap, 446, 839

\bibitem[Guillard et al.(2012a)]{guillard12} Guillard, P., Boulanger, F., Pineau des For{\^e}ts, G., et al.\ 2012, \apj, 749, 158

\bibitem[Guillard et al.(2012b)]{guillard12b} Guillard, P., Ogle, P.~M., Emonts, B.~H.~C., et al.\ 2012, \apj, 747, 95 

\bibitem[Georgiev et al.(2006)]{georgiev06} Georgiev, I.~Y., Hilker, M., Puzia, T.~H., Chanam{\'e}, J., Mieske, S., Goudfrooij, P., Reisenegger, A., \& Infante, L.\ 2006, \aap, 452, 141

\bibitem[Gunn \& Gott(1972)]{gunn72} Gunn, J.~E., \& Gott, J.~R., III 1972, \apj, 176, 1

\bibitem[Hollenbach \& McKee(1989)]{hollenbach89} Hollenbach, D., \& McKee, C.~F.\ 1989, \apj, 342, 306

\bibitem[Houck et al.(2004)]{houck04} Houck, J.~R., et al.\ 2004, \apjs, 154, 18

\bibitem[Iglesias-P{\'a}ramo et al.(2002)]{ip02} Iglesias-P{\'a}ramo, J., Boselli, A., Cortese, L., V{\'{\i}}lchez, J.~M., \& Gavazzi, G.\ 2002, \aap, 384, 383

\bibitem[Ikebe et al.(2002)]{ikebe02} Ikebe, Y., Reiprich, T.~H., B{\"o}hringer, H., Tanaka, Y., \& Kitayama, T.\ 2002, \aap, 383, 773 

\bibitem[Johnstone et al.(2007)]{johnstone07} Johnstone, R.~M., Hatch, N.~A., Ferland, G.~J., et al.\ 2007, \mnras, 382, 1246 

\bibitem[Kaufman et al.(2006)]{kaufman06} Kaufman, M.~J., Wolfire, M.~G., \& Hollenbach, D.~J.\ 2006, \apj, 644, 283

\bibitem[Kenney \& Koopmann(1999)]{kenney99} Kenney, J.~D.~P., \& Koopmann, R.~A.\ 1999, \aj, 117, 181

\bibitem[Kenney et al.(2004)]{kenney04} Kenney, J.~D.~P., van Gorkom, J.~H., \& Vollmer, B.\ 2004, \aj, 127, 3361

\bibitem[Kennicutt \& Kent(1983)]{kennicutt83} Kennicutt, R.~C., Jr., \& Kent, S.~M.\ 1983, \aj, 88, 1094

\bibitem[Koribalski et al.(2004)]{koribalski04} Koribalski, B.~S., et al.\ 2004, \aj, 128, 16 

\bibitem[Le Petit et al.(2006)]{lepetit06} Le Petit, F., Nehm{\'e}, C., Le Bourlot, J., \& Roueff, E.\ 2006, \apjs, 164, 506

\bibitem[Maloney et al.(1996)]{maloney96} Maloney, P.~R., Hollenbach, D.~J., \& Tielens, A.~G.~G.~M.\ 1996, \apj, 466, 561

\bibitem[Mei et al.(2007)]{mei07} Mei, S., Blakeslee, J.~P., C{\^o}t{\'e}, P., et al.\ 2007, \apj, 655, 144 

\bibitem[Mohr et al.(1999)]{mohr99} Mohr, J.~J., Mathiesen, B., \& Evrard, A.~E.\ 1999, \apj, 517, 627

\bibitem[Ogle et al.(2010)]{ogle10} Ogle, P., Boulanger, F., Guillard, P., et al.\ 2010, \apj, 724, 1193

\bibitem[Poggianti et al.(1999)]{poggianti99} Poggianti, B.~M., Smail, I., Dressler, A., et al.\ 1999, \apj, 518, 576

\bibitem[Rieke et al.(2009)]{rieke09} Rieke, G.~H., Alonso-Herrero, A., Weiner, B.~J., et al.\ 2009, \apj, 692, 556

\bibitem[Rigopoulou et al.(2002)]{rigopoulou02} Rigopoulou, D., Kunze, D., Lutz, D., Genzel, R., \& Moorwood, A.~F.~M.\ 2002, \aap, 389, 374 

\bibitem[Roediger \& Br{\"u}ggen(2008)]{roediger08} Roediger, E., \& Br{\"u}ggen, M.\ 2008, \mnras, 388, 465

\bibitem[Roussel et al.(2007)]{roussel07} Roussel, H., et al.\ 2007, \apj, 669, 959 

\bibitem[Rubin(1989)]{rubin89} Rubin, R.~H.\ 1989, \apjs, 69, 897 

\bibitem[Salim et al.(2007)]{salim07} Salim, S., et al.\ 2007, \apjs, 173, 267

\bibitem[Schroeder \& Visvanathan(1996)]{schroeder96} Schroeder, A., \& Visvanathan, N.\ 1996, \aaps, 118, 441

\bibitem[Scott et al.(2010)]{scott10} Scott, T.~C., et al.\ 2010, \mnras, 403, 1175

\bibitem[Sivanandam et al.(2010)]{sivanandam10} Sivanandam, S., Rieke, M.~J., \& Rieke, G.~H.\ 2010, \apj, 717, 147

\bibitem[Smith \& Madden(1997)]{smith97} Smith, B.~J., \& Madden, S.~C.\ 1997, \aj, 114, 138 

\bibitem[Smith et al.(2007)]{smith07a} Smith, J.~D.~T., et al.\ 2007, \pasp, 119, 1133 

\bibitem[Smith et al.(2007)]{smith07b} Smith, J.~D.~T., et al.\ 2007, \apj, 656, 770

\bibitem[Smith et al.(2010)]{smith10} Smith, R.~J., et al.\ 2010, \mnras, 1238 

\bibitem[Sun et al.(2006)]{sun06} Sun, M., Jones, C., Forman, W., Nulsen, P.~E.~J., Donahue, M., \& Voit, G.~M.\ 2006, \apjl, 637, L81 

\bibitem[Sun et al.(2007)]{sun07} Sun, M., Donahue, M., \& Voit, G.~M.\ 2007, \apj, 671, 190

\bibitem[Sun et al.(2010)]{sun10} Sun, M., Donahue, M., Roediger, E., Nulsen, P.~E.~J., Voit, G.~M., Sarazin, C., Forman, W., \& Jones, C.\ 2010, \apj, 708, 946 

\bibitem[Tyler et al.(2013)]{tyler13} Tyler, K.~D., Rieke, G.~H., \& Bai, L.\ 2013, \apj, 773, 86

\bibitem[Vollmer et al.(2006)]{vollmer06} Vollmer, B., Soida, M., Otmianowska-Mazur, K., Kenney, J.~D.~P., van Gorkom, J.~H., \& Beck, R.\ 2006, \aap, 453, 883

\bibitem[Vollmer et al.(2008)]{vollmer08} Vollmer, B., Braine, J., Pappalardo, C., \& Hily-Blant, P.\ 2008, \aap, 491, 455

\bibitem[Vollmer et al.(2012)]{vollmer12} Vollmer, B., Wong, O.~I., Braine, J., Chung, A., \& Kenney, J.~D.~P.\ 2012, \aap, 543, A33

\bibitem[White(2000)]{white00} White, D.~A.\ 2000, \mnras, 312, 663

\bibitem[Wong et al.(2014)]{wong14} Wong, O.~I., Kenney, J.~D.~P., Murphy, E.~J., \& Helou, G.\ 2014, \apj, 783, 109 

\bibitem[Wu et al.(2008)]{wu08} Wu, Y., Bernard-Salas, J., Charmandaris, V., et al.\ 2008, \apj, 673, 193

\bibitem[Yagi et al.(2010)]{yagi10} Yagi, M., Yoshida, M., Komiyama, Y., et al.\ 2010, \aj, 140, 1814 


\end{thebibliography}
\end{document}